\input harvmac
\input epsf
\input amssym
\noblackbox

\newcount\figno

\figno=0
\def\fig#1#2#3{
\par\begingroup\parindent=0pt\leftskip=1cm\rightskip=1cm\parindent=0pt
\baselineskip=11pt
\global\advance\figno by 1
\midinsert
\epsfxsize=#3
\centerline{\epsfbox{#2}}
\vskip 12pt
\centerline{{\bf Figure \the\figno} #1}\par
\endinsert\endgroup\par}
\def\figlabel#1{\xdef#1{\the\figno}}
\def\pano{\par\noindent}

\def\fig#1#2#3{
\par\begingroup\parindent=0pt\leftskip=1cm\rightskip=1cm\parindent=0pt
\baselineskip=11pt
\global\advance\figno by 1
\midinsert
\epsfxsize=#3
\centerline{\epsfbox{#2}}
\vskip 12pt
\centerline{{\bf Figure \the\figno} #1}\par
\endinsert\endgroup\par}
\font\cmss=cmss10
\font\cmsss=cmss10 at 7pt

\def\rlx{\relax\leavevmode}
\def\inbar{\vrule height1.5ex width.4pt depth0pt}
\def\IC{\relax\,\hbox{$\inbar\kern-.3em{\rm C}$}}
\def\IR{\relax{\rm I\kern-.18em R}}
\def\IN{\relax{\rm I\kern-.18em N}}
\def\IP{\relax{\rm I\kern-.18em P}}
\def\IRP{\IR\IP}
\def\ZZ{\rlx\leavevmode\ifmmode\mathchoice{\hbox{\cmss Z\kern-.4em Z}}
 {\hbox{\cmss Z\kern-.4em Z}}{\lower.9pt\hbox{\cmsss Z\kern-.36em Z}}
 {\lower1.2pt\hbox{\cmsss Z\kern-.36em Z}}\else{\cmss Z\kern-.4em Z}\fi}

\def\narrowplus{\kern -.04truein + \kern -.03truein}
\def\narrowminus{- \kern -.04truein}
\def\narrowminussub{\kern -.02truein - \kern -.01truein}

\def\ts{\textstyle}

\def\o#1{\overline{#1}}

\def\sqr#1#2{{\vcenter{\vbox{\hrule height.#2pt
 \hbox{\vrule width.#2pt height#1pt \kern#1pt
 \vrule width.#2pt}\hrule height.#2pt}}}}
\def\square
 {\mathop{\mathchoice{\sqr{12}{15}}{\sqr{9}{12}}{\sqr{6.3}{9}}{\sqr{4.5}{9}}}}


\def\hepth#1{\hbox{hep-th/#1}}

\lref\rgomis{J. Gomis, {\it D-Branes, Holonomy and M-Theory},
Nucl.Phys. B606 (2001) 3, \hepth{0103115}.}

\lref\ratiyah{M. Atiyah and E. Witten, {\it
 M-Theory Dynamics On A Manifold Of $G_2$ Holonomy}, \hepth{0107177}.}

\lref\rwittena{E. Witten, {\it Anomaly Cancellation On Manifolds Of 
 $G_2$ Holonomy}, \hepth{0108165}.}

\lref\rwittenb{E. Witten, {\it Phase Transitions In M-Theory And F-Theory}, 
Nucl. Phys. {\bf B471} (1996) 195, \hepth{9603150}.}

\lref\rbobby{B. Acharya and E. Witten, {\it Chiral Fermions from Manifolds of
    $G_2$ Holonomy}, \hepth{0109152}.}

\lref\ramv{M.~Atiyah, J.~Maldacena, C.~Vafa, 
  {\it An M-theory Flop as a Large N Duality}, 
  \hepth{0011256}.
}

\lref\rvafada{F. Cachazo, K. Intriligator and C. Vafa, {\it 
A Large N Duality via a Geometric Transition}, Nucl.Phys. {\bf B603} (2001) 3,
\hepth{0103067}.
}

\lref\rvafadb{F. Cachazo, S. Katz and C. Vafa, {\it
Geometric Transitions and N=1 Quiver Theories}, 
\hepth{0108120}.
}

\lref\rvafadc{F. Cachazo, B. Fiol, K. Intriligator, S. Katz and C. Vafa,
{\it A Geometric Unification of Dualities}, 
\hepth{0110028}.
}

\lref\raga{M.~Aganagic and C.~Vafa, 
  {\it Mirror Symmetry and a $G_2$ Flop},
  \hepth{0105225}.
}

\lref\ragb{M.~Aganagic and C.~Vafa, 
  {\it $G_2$ Manifolds, Mirror Symmetry and Geometric Engineering},
  \hepth{0110171}.
}

\lref\rvafa{C. Vafa, {\it Superstrings and Topological Strings at Large N},
\hepth{0008142}.}

\lref\rtatar{J.D. Edelstein, K. Oh and R. Tatar, {\it 
Orientifold, Geometric Transition and Large N Duality for SO/Sp Gauge
Theories}, JHEP {\bf 0105} (2001) 009, \hepth{0104037}.} 

\lref\rjoyce{D. Joyce, {\it Compact Manifolds of Special Holonomy},
(Oxford University Press, 2000).}

\lref\rpope{M. Cvetic, G.W. Gibbons, H. Lu and C.N. Pope, {\it 
New Complete Non-compact Spin(7) Manifolds}, \hepth{0103155}.}

\lref\rmoore{J.~A.~Harvey and G.~Moore, 
  {\it Superpotentials and Membrane Instantons}, 
  \hepth{9907026}.
}

\lref\rbrandh{A. Brandhuber, J. Gomis, S.S. Gubser and S. Gukov, {\it
Gauge Theory at Large N and New $G_2$ Holonomy Metrics}, 
Nucl.Phys. {\bf B611} (2001) 179,  \hepth{0106034}.}

\lref\rnunez{Jose D. Edelstein and Carlos Nunez, {\it
D6 branes and M theory geometrical transitions from gauged supergravity},
JHEP {\bf 0104} (2001) 028, \hepth{0103167}.}

\lref\rhern{R. Hernandez, {\it Branes Wrapped on Coassociative Cycles},
\hepth{0106055}.}

\lref\rwald{J.P. Gauntlett, N. Kim, D. Martelli and D. Waldram, {\it
Fivebranes Wrapped on SLAG Three-Cycles and Related Geometry},
\hepth{0110034}.}

\lref\rradu{ K. Dasgupta, K. Oh and R. Tatar, {\it
Geometric Transition, Large N Dualities and MQCD Dynamics},
Nucl. Phys. {\bf B610} (2001) 331, \hepth{0105066};
K. Dasgupta, K. Oh and R. Tatar, {\it Open/Closed String Dualities and 
Seiberg Duality from Geometric  Transitions in M-theory}, \hepth{0106040};
K. Dasgupta, K. Oh, J. Park and R. Tatar, {\it 
Geometric Transition versus Cascading Solution},  \hepth{0110050}.}

\lref\racharya{B. S. Acharya, {\it On Realising N=1 Super Yang-Mills in M
    theory}, \hepth{0011089}.}

\lref\rbwiss{
R.~Blumenhagen and A.~Wi{\ss}kirchen,
  {\it Exactly solvable $(0,2)$ supersymmetric string vacua with GUT
  gauge groups},
  Nucl. Phys. {\bf B454} (1995) 561, \hepth{9506104};
R.~Blumenhagen, R.~Schimmrigk and A.~Wi{\ss}kirchen,
  {\it The $(0,2)$ exactly solvable structure of chiral rings,
  Landau--Ginzburg theories and Calabi--Yau manifolds},
  Nucl.~Phys.~{\bf B461} (1996) 460, \hepth{9510055};
R.~Blumenhagen, R.~Schimmrigk and A.~Wi{\ss}kirchen,
  {\it $(0,2)$ Mirror Symmetry}, Nucl.~Phys.~{\bf B486} (1997) 598, 
  \hepth{9609167}.
}

\lref\rsy{A.~N.~Schellekens and S.~Yankielowicz,
  {\it New modular invariants for $N=2$ tensor products and
  four-dimensional strings},
  Nucl.~Phys.~{\bf B330} (1990) 103.
}

\lref\rbanks{T.~Banks, L.~J.~Dixon, D.~Friedan and E.~ Martinec,
  {\it Phenomenology and conformal field theory, 
  or Can string theory predict the weak mixing angle?}, 
  Nucl.~Phys.~{\bf B299} (1988) 613.
}

\lref\rgepner{D.~Gepner,
  {\it Space--time supersymmetry in compactified string theory and
  superconformal models},
  Nucl.~Phys.~{\bf B296} (1988) 757.
}

\lref\rgq{D.~Gepner and Z.~Qiu,
  {\it Modular invariant partition functions for para\-fermionic field
  theories},
  Nucl.~Phys.~{\bf B285} (1987) 423.
}

\lref\rlgcy{D.~Gepner,
  {\it Exactly solvable string compactification on manifolds of $SU(n)$
  holonomy},
  Phys.~Lett.~{\bf B199} (1987) 380.
}

\lref\reot{T.~Eguchi, H.~Ooguri, A.~Taormina and S.~K.~Yang,
  {\it Superconformal algebras and string compactification on manifolds
  with $SU(n)$ holonomy}, 
  Nucl.~Phys.~{\bf B315} (1989) 193.
}

\lref\rtensorb{
J.~Fuchs, A.~Klemm, C.~Scheich and M.~Schmidt,
  {\it Gepner models with arbitrary invariants and the associated
  Calabi--Yau spaces},
  Phys.~Lett.~{\bf B232} (1989) 317;
J.~Fuchs, A.~Klemm, C.~Scheich and M.~Schmidt,
  {\it Spectra and symmetries of Gepner models compared to Calabi--Yau
  compactifications},
  Ann.~Phys.~{\bf 204} (1990) 1.
}

\lref\rschimmrigk{M. Lynker and R. Schimmrigk, {\it On the Spectrum of
(2,2) Compactification of the Heterotic String on Conformal Fields Theories}
Phys. Lett. {\bf B215} (1988) 681;
M. Lynker and R. Schimmrigk, {\it A-D-E Quantum Calabi-Yau Manifolds},
Nucl. Phys. {\bf B339} (1990) 121.}

\lref\rdbranes{I.~Brunner, M.~R.~Douglas, A.~Lawrence and
  C.~Romelsberger, 
  {\it D-branes on the Quintic}, 
  JHEP~{\bf 0008} (2000) 015,
  \hepth{9906200}.
}

\lref\rkdbranes{I.~Brunner and  J.~Distler, 
  {\it Torsion D-Branes in Nongeometrical Phases},
  \hepth{0102018}.
}

\lref\rbound{A. Recknagel and V. Schomerus, 
  {\it D-branes in Gepner models},
  Nucl.~Phys.~{\bf B531} (1998) 185, 
  \hepth{9712186}.
}

\lref\rzamo{A.B. Zamolodchikov and V.A. Fateev, {\it Nonlocal (parafermion)
currents in two dimensional conformal quantum field theory and self-dual
critical points in $\ZZ_N$-symmetric statistical systems},
Zh. Eksp. Teor. Fiz. {\bf 89} (1985) 380.}

\lref\rcandelas{P. Candelas, X.C. De La Ossa, P.S. Green and L. Parkes, 
{\it A Pair of Calabi--Yau Manifolds as an Exactly Soluable Superconformal 
Theory}, Nucl. Phys. {\bf B359} (1991) 21-74.} 

\lref\raspinwall{P.~Aspinwall,
  {\it The Moduli Space of N = 2 Superconformal Field Theories}, 
  \hepth{9412115}.
}

\lref\rschell{
A.~N.~Schellekens and S.~Yankielowicz, 
  {\it Extended chiral algebras and modular invariant 
  partition functions},
  Nucl.~Phys.~{\bf B327} (1989) 673; 
A.~N.~Schellekens and S.~Yankielowicz, 
  {\it Simple currents, modular invariants and Fixed Points},
  Int.~J.~Mod.~Phys.~{\bf A5} (1990) 2903.
}

\lref\rvascha{S.~L.~Shatashvili and C.~Vafa, 
  {\it  Superstrings and Manifolds of Exceptional Holonomy}, 
  \hepth{9407025}.
}

\lref\rjose{J.~M.~Figueroa-O'Farrill, 
  {\it A note on the extended superconformal
  algebras associated with manifolds of exceptional holonomy}, 
  Phys.~Lett.~{\bf B392} (1997) 77, 
  \hepth{9609113}.
}

\lref\rsugi{K.~Sugiyama and  S.~Yamaguchi, 
  {\it Cascade of Special Holonomy
  Manifolds and Heterotic String Theory},  
  \hepth{0108219}.
}

\lref\reguchi{T.~Eguchi and  Y.~Sugawara, 
  {\it CFT Description of String
  Theory Compactified on Non-compact Manifolds with $G_2$ Holonomy}, 
  \hepth{0108091}.
}

\lref\rkachru{S.~Kachru and J.~McGreevy, 
  {\it M-theory on Manifolds of $G_2$
  Holonomy and Type IIA Orientifolds}, 
  JHEP~{\bf 0106} (2001) 027, 
  \hepth{0103223}.
}

\lref\rcvetic{
M.~Cvetic, G.~Shiu and A.~M.~Uranga, 
  {\it Three--Family Supersymmetric Standard-like Models 
  from Intersecting Brane Worlds}, 
  \hepth{0107143}; 
M.~Cvetic, G.~Shiu and A.~M.~Uranga,
  {\it Chiral Four-Dimensional N=1 Supersymmetric Type IIA 
  Orientifolds from Intersecting D6-Branes}, 
  \hepth{0107166}.
}

\lref\rbgkl{R.~Blumenhagen, L.~G\"orlich, B.~K\"ors and D.~L\"ust, 
  {\it Noncommutative Compactifications of Type I Strings on 
  Tori with Magnetic Background Flux}, 
  JHEP~{\bf 0010} (2000) 006, 
  \hepth{0007024};
R.~Blumenhagen, B.~K\"ors and D.~L\"ust, 
  {\it Type I Strings with $F$ and $B$-Flux}, 
  JHEP~{\bf 0102} (2001) 030,
  \hepth{0012156};
R.~Blumenhagen, L.~G\"orlich and B.~K\"ors,
  {\it Supersymmetric 4D Orientifolds of Type IIA with D6-branes at Angles},
  JHEP~{\bf 0001} (2000) 040, 
  \hepth{9912204}. 
}

\lref\rpioline{H.~Partouche and B.~Pioline, 
  {\it Rolling among $G_2$ vacua}, 
  JHEP~{\bf 0103} (2001) 005, 
  \hepth{0011130}.
}
 
\lref\rkehagiasa{P.~Kaste, A.~Kehagias and H.~Partouche, 
  {\it Phases of supersymmetric gauge theories from M-theory 
  on $G_2$ manifolds},
  JHEP~{\bf 0105} (2001) 058, 
  \hepth{0104124}.
}

\lref\rkehagiasb{A.~Giveon, A.~Kehagias and H.~Partouche, 
  {\it Geometric Transitions, Brane Dynamics and Gauge Theories}, 
  \hepth{0110115}.
}

\lref\rhornfeck{C.~Hornfeck, 
  {\it W-algebras with set of primary fields of
  dimensions (3, 4, 5) and (3,4,5,6)}, 
  Nucl.~Phys.~{\bf B407} (1993) 237, 
  \hepth{9212104}.
}

\lref\runify{
R.~Blumenhagen, W.~Eholzer, A.~Honecker, K.~Hornfeck and R.~H\"ubel, 
  {\it Coset Realization of Unifying W-Algebras},
  Int.~J.~Mod.~Phys.~{\bf A10} (1995) 2367, 
  \hepth{9406203};
R.~Blumenhagen, W.~Eholzer, A.~Honecker, K.~Hornfeck and R.~H\"ubel, 
  {\it Unifying W-Algebras}, 
  Phys.~Lett.~{\bf B332} (1994) 51, 
  \hepth{9404113}.
}

\lref\reholzer{W.~Eholzer and R.~H\"ubel, 
  {\it Fusion Algebras of Fermionic Rational Conformal Field 
  Theories via a Generalized Verlinde Formula},
  Nucl.~Phys.~{\bf B414} (1994) 348, 
  \break
  \hepth{9307031}.
}

\lref\ralt{D. Altschuler, {\it Quantum equivalence of coset space models},
Nucl.~Phys.~{\bf B313} (1989) 293.}

\lref\rsagnotti{M. Bianchi, G. Pradisi and A. Sagnotti, {\it 
Toroidal Compactification and Symmetry Breaking in Open String Theories}
Nucl. Phys. {\bf B376} (1992) 365.}

\Title{\vbox{
 \hbox{HU--EP--01/47}
 \hbox{hep-th/0110232}}}
{\vbox{\centerline{Superconformal Field Theories for} 
\vskip 0.4cm
      {\centerline{Compact G$_2$ Manifolds} }
}}
\centerline{Ralph Blumenhagen \footnote{$^1$}{{\tt e-mail:
 blumenha@physik.hu-berlin.de}} and 
Volker Braun \footnote{$^2$}{{\tt e-mail: volker.braun@physik.hu-berlin.de}}
} 
\bigskip
\centerline{\it Humboldt-Universit\"at zu Berlin, Institut f\"ur  
Physik,}
\centerline{\it Invalidenstrasse 110, 10115 Berlin, Germany}
\smallskip
\bigskip
\centerline{\bf Abstract}
\noindent
We present the construction of exactly solvable superconformal
field theories describing Type II string models compactified 
on compact $G_2$ manifolds. 
These models are defined  by anti-holomorphic quotients of  
the form (CY$\times S^1)/\ZZ_2$, where we realize the Calabi-Yau as 
a Gepner model.
In the superconformal field theory the $\ZZ_2$ acts as
charge conjugation implying  that the representation theory
of a ${\cal W}(2,4,6,8,10)$ algebra plays an important r\^ole
in the construction of these models.  
Intriguingly, in all three examples we study, including the quintic, 
the massless spectrum in the  $\ZZ_2$ twisted sector 
of the superconformal field theory  differs from what one expects
from the supergravity computation. This discrepancy is explained
by the presence of a discrete  NS-NS background two-form flux in the 
Gepner model.

\bigskip

\Date{10/2001}
\newsec{Introduction}

Recently, we have seen an intensified effort to reveal 
the structure of M-theory compactifications on manifolds with
exceptional holonomy \refs{\rjoyce\rmoore\racharya\rpioline\rgomis
\rkachru\rkehagiasa
\ragb\ratiyah\rwittena\rbobby\rkehagiasb-\raga}. 
The main focus in recent developments
was on non-compact examples of $G_2$ and Spin$(7)$ holonomy
where explicit metrics have been constructed 
\refs{\rpope\rnunez\rbrandh\rnunez\rhern-\rwald}. Moreover,
flop transitions on such manifolds allow for  purely geometric M-theory lifts 
of the so-called Vafa duality \refs{\rvafa\ramv\rvafada\rtatar\rvafada
\rvafadb-\rvafadc}. An alternative lift of Vafa duality to M-theory 
has been considered in \rradu.

M-theory on compact seven dimensional $G_2$ manifolds  are of special 
interest, as they lead to  four
dimensional effective theories with ${\cal N}=1$ space-time
supersymmetry. Since M-theory on smooth $G_2$ manifolds only gives
rise to abelian gauge symmetries with non-chiral matter, it is clear
that interesting phenomenology can only
be realized on singular spaces. The kinds of singularities giving
rise to non-abelian gauge symmetries and chiral matter have
been analyzed in \refs{\ratiyah,\rwittena,\rbobby}. 

Most explicit compact $G_2$ manifolds
constructed so far 
are given by certain toroidal orbifolds. This class includes both 
the models  constructed by Joyce
and  the ones resulting from an M-theory lift of certain Type IIA orientifolds
with D6-branes \refs{\rcvetic,\rbgkl}. 
Another large class of $G_2$ manifolds is supposed to result
from anti-holomorphic $\ZZ_2$ quotients of Calabi-Yau manifolds
times a circle. Phase transitions in the M-theory moduli space
of such manifolds have been investigated in 
\refs{\rpioline,\rkehagiasa,\rkehagiasb}.

Since one is  not equipped with a microscopic quantum M-theory, 
one can only study such models in the large radius limit
where the supergravity approximation is valid. What one can do however is 
to compactify M-theory on a further $S^1$
down to three dimensions and employ the duality with Type IIA
string theory, where computations in the small distance
regime are in principle possible. 
For carrying out  such a computation it is necessary to exactly solve the 
non-linear
sigma model in  this curved background. Even though, except  for toroidal
orbifolds, this is technically beyond our abilities, sometimes
pure conformal field theory (CFT) considerations have proven to be successful
in providing models which accidentally correspond to certain points
in the deep interior of, for instance, the Calabi-Yau moduli space. 
The most prominent examples are  certainly
given by the so-called Gepner models \refs{\rgepner,\rlgcy,\reot} , 
which use tensor products
of minimal models of the ${\cal N}=2$ super Virasoro
algebra  equipped by  a GSO projection in the internal conformal 
field theory.  
Such superconformal field theories (SCFT) have been identified
with certain points in the moduli space of Calabi-Yau threefolds
given by Fermat type hypersurfaces in weighted projective spaces 
\refs{\rlgcy,\rschimmrigk,\rtensorb}. 
Other examples are given by so called $(0,2)$ generalizations
of the Gepner models which appear in heterotic compactifications
with ${\cal N}=1$ supersymmetry in four dimensions \rbwiss. 
Moreover, Gepner models also served as a powerful tool
in investigating stable BPS as well as non-BPS D-branes present in Calabi-Yau 
compactifications in the stringy regime \refs{\rbound,\rdbranes,\rkdbranes}. 

In this paper we present a class of exactly solvable superconformal
fields theories which are argued to correspond to certain points in the
moduli space of Type II compactifications on $G_2$ manifolds. These
manifolds are given by anti-holomorphic quotients of the form
(CY$\times S^1)/\ZZ_2$, where the Calabi-Yau manifold is given by a
Fermat type hypersurface in a weighted projective space and is described
in the SCFT by a Gepner model.
Note that the general structure of the SCFT describing $G_2$ manifolds
has been investigated in \refs{\rvascha,\rjose,\rsugi} but except
toroidal orbifolds no explicit SCFT has been found so far.
On the technical level we have to implement the anti-holomorphic
$\ZZ_2$ action in the corresponding Gepner model, which turns out to be 
nothing else
than conjugation of the $U(1)$ charges in each factor theory. 
However, the determination of the action of  charge conjugation
on all states in the Hilbert space of a  Gepner model is quite challenging and 
we present here the solution to this problem at least  for the 
$k\in\{1,2,3,6\}$ minimal models. We argue that the general
solution to this problem is related to the so far unknown
representation theory of a ${\cal W}(2,4,6,8,10)$ algebra.

Once the $\ZZ_2$ action is known on the entire Hilbert space, it is
fairly straightforward to compute the orbifold partition
function, including the new $\ZZ_2$ twisted sector, and to determine
the massless spectrum in three space-time dimensions.
For all models studied in this paper the massless spectrum disagrees
with what one naively expects from the supergravity analysis. 
However, we will show that this is not surprising at all. 
It is known from the mirror symmetry analysis that
 the Gepner models correspond to points in Calabi-Yau
moduli space with radii at the string scale and  non-trivial
background NS-NS two-form fluxes turned on. Under the anti-holomorphic 
involution the continuous moduli related to these fluxes are projected out,
but nevertheless certain discrete values are still allowed. Therefore, the
resolution of the puzzle stated  above is simply that the supergravity
model and the Gepner model occupy  disconnected branches
of the $G_2$ moduli space. Besides that, one expects world-sheet 
instanton corrections to be relevant in the stringy regime anyway
which might lead to different phases with different massless modes.

This paper is organized as follows. In section 2 we briefly
review the relevant aspects of the construction of Gepner models. 
In section 3 we determine the action of the anti-holomorphic 
involution for some of the ${\cal N}=2$ unitary models, which allows
us to study at least some of the Gepner models. 
In section 4, for the quotient of the quintic 
$\IP_5[5]$,  we explicitly compute the one-loop partition 
function and determine the massless spectrum including fields from
the $\ZZ_2$ twisted sector. In section 5  we derive
the expected geometric large radius result for the quintic and 
point out the discrepancy with the SCFT result. Section 6 
provides more complicated SCFT examples involving 
also the $(k=6)$ unitary model. In section 7 we present the computation
of these $G_2$ compactifications in the supergravity limit and compare
the results to the SCFT models. In addition we provide some
material on the Cartan-Leray spectral sequence. Section 8  
provides the resolution of the puzzle concerning the different
results for the SCFT and supergravity computation.
Finally, section 9 contains our conclusions and mentions some open problems.

\newsec{Review of Gepner models}

The goal of this paper is to study Type II compactifications
on $G_2$ manifolds of the form 
\eqn\form{   {{\rm CY}\,\times S^1\over \sigma^* } ,}
where the Calabi-Yau threefold is given by 
a Fermat type hypersurface in a weighted projective space 
$\IP_{w_1,w_2,w_3,w_4,w_5}[d]$ with $d=\sum w_i$.
The anti-holomorphic involution acts on the homogeneous
coordinates $z_i$ of the projective space by complex conjugation and
on the real coordinate, $y$, parameterizing  the $S^1$ by a reflection. 
Going deep inside the K\"ahler moduli space of the Calabi-Yau
there exists a point where the exact ${\cal N}=(2,2)$ superconformal
field theory is  explicitly known and described by a Gepner model \rgepner. 
In the following we work at this special point in moduli space. 
However, before continuing the construction of SCFTs for
the $G_2$ manifolds \form, we need to review some aspects
of Gepner's construction. For readers not familiar with Gepner models,
we would like to refer them to the original literature \rgepner.

\subsec{N=2 unitary models}

Gepner models are given by tensor products of rational models
of the ${\cal N}=2$ super Virasoro algebra, so that the central
charge of all tensor models adds up to $c=9$.
The ${\cal N}=2$ extension of the Virasoro algebra contains besides
the energy momentum tensor $L$ two fermionic superpartners
$G^\pm$ of conformal dimension $h={3\over 2}$ and
one bosonic $U(1)$ current $j$ of conformal dimension $h=1$.
For future reference we give here the explicit form of
the  ${\cal N}=2$ super Virasoro algebra
\eqn\vir{ \eqalign{
        [L_m,L_n]&=(m-n)\, L_{m+n} + {c\over 12}\, n(n^2-1)\,
        \delta_{m+n,0}\cr
        [L_m,j_n]&=-n\,  j_{m+n} \cr
        [L_m,G^\pm_r]&=\left( {m\over 2}-r\right) \,  G^\pm_{m+r} \cr
        [j_m,j_n]&={c\over 3}\, n \delta_{m+n,0} \cr 
       [j_m,G^\pm_r]&=\pm G^\pm_{m+r} \cr
       \{G^+_r,G^-_s\}&=2L_{r+s} + (r-s)\, j_{r+s}+
           {c\over 3}\, \left( r^2-{1\over 4} \right) 
        \delta_{r+s,0}\cr
       \{G^+_r,G^+_s\} &= \{G^-_r,G^-_s\}=0. \cr }}
The rational models are classified and the central charge is known to be 
restricted to  the discrete series 
\eqn\disse{  c={3k\over k+2},\quad k\in\ZZ_+ .}
For each level $k$ there exists only a finite number of highest weight 
representations $(h,q)$ labeled by their conformal dimension and their
$U(1)$ charge
\eqn\highest{\eqalign{
           h&={l(l+2)-m^2\over 4(k+2)} +{s^2\over 8} \cr
           q&=-{m\over k+2}+{s\over 2}. \cr}}
The three indices $(l,m,s)$ are restricted to lie in the standard range
\eqn\standard{\eqalign{
          &0\le l\le k, \cr
          &0\le |m-s| \le l, \cr
           & s=\cases{ 0,2 & NS sector \cr
                       \pm 1 & R sector \cr} \cr
           & m=m\ {\rm mod}\ 2(k+2), \ \ s=s\ {\rm mod}\ 4,\ \  
                l+m+s=0\ {\rm mod}\, 2.\cr }}
Note that all superconformal characters have  been split
into two bosonic pieces like for instance in the NS-sector
\eqn\split{ 
          \chi^l_m=\chi^l_{m,s=0}+\chi^l_{m,s=2} .}
These unitary models can be written as the product of the parafermions
\refs{\rzamo,\rgq} and a free $U(1)$ current
\eqn\coset{     {SU(2)_k\over U(1) }\times U(1), }
where the quotient of the $SU(2)_k$ affine Lie algebra
by the Cartan $U(1)$ is precisely the parafermionic CFT and
the additional $U(1)$ can be identified with the abelian current in the
${\cal N}=2$ Virasoro algebra. 
The realization as a coset \coset\ enables one to easily
determine the characters of the unitary  representations
of the ${\cal N}=2$ Virasoro algebra as branching functions
\refs{\rgq,\rgepner}
\eqn\char{
    \chi^l_{m,s}(\tau)=\sum_{j=1}^k  C^l_{m-(4j+s)}(\tau)\
              \Theta_{2m-(k+2)(4j+s),2k(k+2)}\left(\tau,{z\over k+2}\right) .}
The $C^l_m$ are called string functions of $SU(2)_k$ and are related to
the characters $\kappa^l_m$ of the parafermionic theory by
\eqn\charpar{
           \kappa^l_m(\tau)=\eta(\tau)\, C^l_{m}(\tau) .}
These string functions are explicitly known 
\eqn\stringf{
  C^l_m(\tau)=\eta(\tau)^{-3}\hskip -2cm \sum_{  
  \matrix{ \scriptstyle (x,y)\in \IR^2 \cr
           \scriptstyle -|x|<|y|\le |x| \cr
           \scriptstyle
      (x,y)\ {\rm or}\ (1/2-x,1/2+y)\in ((l+1)/2(k+2),m/2k)+\ZZ^2 \cr}}
  \hskip -2cm     {\rm sign}(x)\, q^{(k+2)x^2-ky^2}}
with $q=e^{2\pi i\tau}$ and will play an important r\^ole in the following. 

\subsec{Gepner's construction}

It had been known that in order to get ${\cal N}=2$ space-time 
supersymmetry in a four dimensional Type II compactification
one needs ${\cal N}=(2,2)$ supersymmetry on the two-dimensional
world-sheet \rbanks. Moreover, there must exist a spectral flow operator
relating the NS and the R sector of the SCFT. 

Gepner realized the internal SCFT describing the Calabi-Yau manifold
by tensor products of unitary models of the ${\cal N}=2$ Virasoro algebra
such that the central charges adds up to $c=9$. 
The remaining central charge of $c=3$ is occupied  in light-cone
gauge by two flat bosons $X^\mu$ with $\mu=2,3$ and their fermionic
superpartners
$\psi^\mu$. The latter ones yield  a realization of the $SO(2)_1$
current algebra, namely in the NS sector the vacuum $(O_2)_{h=0,q=0}$ and the 
vector $(V_2)_{h=1/2,q=1}$ representation and in the R sector the spinor
$(S_2)_{h=1/8,q=1/2}$ and the antispinor $(C_2)_{h=1/8,q=-1/2}$ 
representation. The characters for these representation read
\eqn\charrep{\eqalign{
  O_2&={1\over 2}\left( {\theta_3\over \eta} +{\theta_4\over \eta}\right) \cr
  V_2&={1\over 2}\left( {\theta_3\over \eta} -{\theta_4\over \eta}\right) \cr
    S_2&={1\over 2}\left( {\theta_2\over \eta }\right) \cr
    C_2&={1\over 2}\left( {\theta_2\over \eta} \right). \cr}}
Neglecting in the following the flat space-time bosons, the starting point
for the Gepner construction is the tensor product
\eqn\geptens{
     \bigotimes_{i=1}^N (k_i)\times SO(2)_1 }
which contains highest weights denoted by
\eqn\hwgep{ 
        \prod_{i=1}^N (l_i,m_i,s_i)\times (\phi) }
with $\phi\in\{O_2,V_2,S_2,C_2\}$.

In order to get a space-time supersymmetric string theory one has
to implement a GSO projection, which in this case projects 
onto states with odd overall $U(1)$ charge both in the left moving
and the right moving sector. 
More formally, the GSO projection is realized by constructing
a new modular invariant partition function  utilizing the 
simple current \footnote{$^1$}
{More precisely, one first has to apply the bosonic string map exchanging
$SO(2)$ with $SO(10)\times E_8$, apply the simple current
techniques and finally map  back to $SO(2)_1$.} \refs{\rschell,\rsy}
\eqn\simple{ 
        J_{GSO}=(0,1,1)^N\otimes (C_2) . }
The effect of this simple current construction is that it projects
onto states with odd overall $U(1)$ charge and arranges the
surviving fields into orbits of finite length under the action of $J_{GSO}$. 
Besides the GSO projection one also has to make sure that from
the individual factor theories only NS respectively R  sector states are 
combined. This projection is implemented by the simple currents
\eqn\simpleb{ 
        J_{a}=\prod_{i=1}^{a-1} (0,0,0)\otimes (0,0,2)
               \otimes \prod_{i=a+1}^{N} (0,0,0)  \otimes (V_2)  }
with $a=1,\ldots,N$.
All these projections imply that in the vacuum orbit 
the massless states
\eqn\gravi{\eqalign{ &\left[ (0,0,0)^N\otimes (V_2) \right]_L \times
                     \left[ (0,1,1)^N\otimes (C_2) \right]_R, \cr
                     &\left[ (0,1,1)^N\otimes (C_2) \right]_L\times
                    \left[ (0,0,0)^N\otimes (V_2) \right]_R   \cr }}
and their charge conjugates survive, which are precisely 
the gravitinos of ${\cal N}=2$ space-time supersymmetry in four 
dimensions. 

The massless spectrum of such a Gepner model can be determined
from the modular invariant partition function. 
The vacuum orbit gives rise to the ${\cal N}=2$ supergravity multiplet
in addition to one hypermultiplet containing the dilaton and
the dualized NS-NS two form.
Chiral states  of the form
\eqn\chiral{ \left[ (h=1/2,q=1)\otimes (O_2) \right]_L \times
                     \left[ (h=1/2,q=1) \otimes (O_2) \right]_R }
and their charge conjugates give rise to one hyper(vector)-multiplet, 
whereas anti-chiral states
\eqn\antichiral{ \left[ (h=1/2,q=1)\otimes (O_2) \right]_L \times
                     \left[ (h=1/2,q=-1) \otimes (O_2) \right]_R }
and their charge conjugates give rise to one vector(hyper)-multiplet in
Type IIA(IIB) string theory.

\newsec{Anti-holomorphic involution}

After we have reviewed the main ingredients of  Gepner models in the
last section we now move forward to the construction of SCFTs
for $G_2$ manifolds. Starting with a Gepner model we also
compactify the  transversal boson $X^3$   on a circle of radius
$R$ so that in light-cone gauge we are left with only one non-compact
direction $X^2$. 
The next step is to realize the anti-holomorphic involution $\sigma^*$ 
in the Gepner model SCFT. Since formally the homogeneous coordinates
$Z_i$ can be identified with the chiral fields $(l_i,m_i,s_i)=(1,1,0)$ and
the complex conjugates $\o{Z}_i$ with $(l_i,m_i,s_i)=(1,-1,0)$,
a natural ansatz is that charge conjugation in each individual
factor theory of the SCFT is equivalent to complex conjugation 
in the corresponding geometry.  

In the remainder of this section  we study the action of this charge 
conjugation
on states in the Hilbert space of one individual ${\cal N}=2$ unitary model. 
On the level of the ${\cal N}=2$ super Virasoro algebra \vir\
charge conjugation acts as 
\eqn\chragecomj{ L_m\to L_m,\quad j_m\to -j_m,\quad 
               G^+_r \leftrightarrow G^-_r }
and is an automorphism of the algebra. The even generators are 
$L$ and ${1\over \sqrt{2}}(G^++G^-)$ which form the ${\cal N}=1$ superconformal
algebra.  Thus generically only an ${\cal N}=1$ superconformal
symmetry survives the $\ZZ_2$ projection. 
Apparently,  the states in the highest weight representation (HWR)
$(l,m,s)$ are mapped to states in the HWR
$(l,-m,-s)$. Thus as long as $m\ne 0$ or $s\ne 0$ two states
are simply exchanged under the action of $\sigma^*$. In particular,
on all states in the R sector $\sigma^*$ acts by exchange of two states
\footnote{$^1$}{Actually, for $k$ even $(l,m,s)=(l,(k+2)/2,1)$ are uncharged 
HWR in the R-sector, but since they have to combine with the spinor or
antispinor representation of $SO(2)$ they are all projected
out by the GSO projection.}.  
However, for states in HWRs of the form $(l,0,0)$ the situation
gets more complicated. Remember the form of the character 
in such a representation
\eqn\charc{
    \chi^l_{0,0}(\tau)=\sum_{j=1}^k  C^l_{-4j}(\tau)\
              \Theta_{-4(k+2)j,2k(k+2)}\left(\tau,{z\over k+2}\right) .}
and that charge conjugation will map
$\Theta_{m,2k(k+2)}$ to $\Theta_{-m,2k(k+2)}$.
Therefore the involution might act non-trivially 
only on states contained in 
\eqn\chard{ C^l_0(\tau)\, \Theta_{0,2k(k+2)}(\tau).}
By the same argument as before only the single uncharged ground state 
in $\Theta_{0,2k(k+2)}$ is not mapped to a different 
 state with opposite charge. 
Thus we conclude that $\sigma^*$ can only act non-trivially on those
states in the HWR $(l,0,0)$ which are counted
in $C^l_0$. As expected these are precisely the neutral states 
in the HWR $(l,0,0)$. Since $j$ is a free field, we can always factor
out its contribution, which is precisely given by the Dedekind 
$\eta$-function
\eqn\dede{ C^l_0(\tau)={1\over \eta(\tau)}\kappa^l_0(\tau) }
where $\kappa^l_0$ are characters of HWRs of the parafermions. 
Unfortunately, the determination of the action of the anti-holomorphic
involution on the neutral states contained in $\kappa^l_0$ turns
out to be a highly non-trivial task. But in a case by case study,
we have managed to find a satisfactory solution at least for the four 
unitary models $k\in\{1,2,3,6\}$. In the following we present
our analysis for these four cases separately.
 
\subsec{Action of $\sigma^*$ for the (k=1) model}

In this case the central charge is $c=1$ so that the parafermionic
part is actually trivial, $\kappa^0_0=1$. 
The only interesting HWR is $(l,m,s)=(0,0,0)$.
The fact that the parafermionic part is trivial means that all
neutral states in this representation are generated by $j_m$ alone. 
Said differently, there are so many null-states in this Verma module
that all states containing modes $L_m$, $G^\pm_r$ can be expressed in terms
of the modes $j_m$. 
The action of $\sigma^*$ on $j_m$ is  known  \chragecomj, so that
for the trace over  the HWR $(l,m,s)=(0,0,0)$ with an $\sigma^*$ 
insertion we get 
\eqn\sigmaeins{  \chi^0_{0,0}(\sigma^*)={\rm Tr}_{{\cal H}^0_{0,0}}\left(
                          \sigma^* e^{2\pi i \tau L_0} \right)= 
                  \sqrt{ 2\eta\over \theta_2 }  .}

\subsec{Action of $\sigma^*$ for the (k=2) model}

In this case we have two interesting HWRs, $(l,m,s)=(0,0,0)$ and 
$(l,m,s)=(2,0,0)$. The central charge is $c={3\over 2}$ so that
the parafermions contribute $c={1\over 2}$. It is a well known 
fact that the first non-trivial parafermionic model is identical
to the Ising model, which can be considered as  the first unitary 
model of the ${\cal N}=0$ Virasoro algebra. For the general parafermionic
theory there exists a duality of coset models \ralt\
\eqn\dual{    {SU(2)_k\over U(1)}={ SU(k)_1\times SU(k)_1\over SU(k)_2 }, }
which means that the $k$th parafermionic model is  identical
to the  first unitary model of the ${\cal WA}_{k-1}$ algebra.
This ${\cal W}$ algebra is generated by primary fields of conformal
dimension $\Delta\in \{2,3,\ldots,k\}$. Thus, naively one might expect
that the chiral symmetry algebra of the parafermions  contains infinitely
many generators. However, as shown in \runify\ this is actually not true. 
All ${\cal WA}_k$ algebras truncate for the first unitary model, i.e.
for $c={2(k-1)\over k+2}$, 
to a ${\cal W}(2,3,4,5)$\footnote{$^1$}{We denote a ${\cal W}$-algebra
with generators of conformal dimension $h\in\{2,\Delta_1,\ldots,\Delta_n\}$ as
${\cal W}(2,\Delta_1,\ldots,\Delta_n)$.}
algebra, which is different from the 
${\cal WA}_4$ algebra. 
Moreover, this ${\cal W}(2,3,4,5)$ algebra truncates for the first
three parafermionic models $k=2,3,4$ 
to the algebras ${\cal W}(2)$, ${\cal W}(2,3)$ and
${\cal W}(2,3,4)$, respectively. Please consult reference \runify\ for
more details. 

Back to the $(k=2)$ model, from the fact that the parafermionic model
is contained in the unitary series of the Virasoro algebra we conclude
that there is still a sufficient number of 
null-states in the Verma module that all uncharged states
generated by $G^\pm_r$ can be expressed by $j_m$ and $L_m$.
Using that the $c={1\over 2}$ theory is given by one free world-sheet
fermion with its different spin structures we can write
\eqn\sigmazwei{\eqalign{  \chi^0_{0,0}(\sigma^*)&= 
                  \sqrt{ 2\eta\over \theta_2 }\, {1\over 2}
               \left(\sqrt{\theta_3\over \eta}+\sqrt{\theta_4\over
                   \eta}\right)\cr
           \chi^2_{0,0}(\sigma^*)&= 
                  \sqrt{ 2\eta\over \theta_2 }\, {1\over 2}
               \left(\sqrt{\theta_3\over \eta}-\sqrt{\theta_4\over
                   \eta}\right).\cr }}

\subsec{Action of $\sigma^*$ for the (k=3) model}

Slowly the situation becomes more complicated. In this case the central
charge is $c={9\over 5}$ where the parafermions contribute 
$c={4\over 5}$. This value for $c$ is both contained in the unitary series
of the Virasoro algebra and in the unitary series of the ${\cal WA}_2$
algebra and the model is known as the 3-states Potts model. 
The parafermionic respectively  ${\cal WA}_2$ 
characters can be written in terms of characters of the $k=5$
Virasoro unitary model
\eqn\potts{  \kappa^0_0=\chi_0+ \chi_3,\quad\quad 
                \kappa^2_0=\chi_{2\over 5}+ \chi_{7\over 5} ,}
where the indices  denote the conformal dimensions of the HWR. 
Please consult appendix A for some basis data of the representation
theory of the Virasoro algebra. The interpretation of $\kappa^0_0$ 
in terms of the generators of the ${\cal N}=2$ Virasoro
algebra is as follows. Now, not the entire Hilbert space is generated by
$j_m$ and $L_m$ alone. Instead at conformal dimension three a new field, $W_3$,
appears. However, the normal ordered product of this field with itself 
gives rise to a null-state. The precise form of this field in terms
of the generators $j,L,G^\pm$ can be found in \runify. 
What is important for us is that the new mode at level three
can only be $G^+_{-{3\over 2}}\, G^-_{-{3\over 2}}|0\rangle$.
Since the $G^\pm$ anticommute, under charge conjugation  this state
picks up a minus sign so that the action of $\sigma^*$ on the 
HWRs $(l,m,s)=(0,0,0)$ and $(l,m,s)=(2,0,0)$ reads 
\eqn\sigmadrei{\eqalign{  \chi^0_{0,0}(\sigma^*)&= 
                  \sqrt{ 2\eta\over \theta_2 }\, \left(\chi_0-\chi_3\right)\cr
           \chi^2_{0,0}(\sigma^*)&= 
                  \sqrt{ 2\eta\over \theta_2 }\, 
                  \left(\chi_{2\over 5}-\chi_{7\over 5}\right).\cr}}
Note that this result is not obvious from the very beginning and we really
needed to perform a quite detailed analysis of the uncharged states. 

That we found such a simple
answer was only possible due to the decomposition of the parafermionic
character in terms of Virasoro characters.
The results presented in section 4 will provide a highly 
nontrivial consistency check for the correctness of \sigmadrei.
In particular, after a modular $S$
transformation of \sigmadrei\ one obtains the $\sigma^*$ twisted sector.
This sector  first must allow the interpretation as a partition function,
second must satisfy level matching and third  must be free of  any tachyonic
states.

For $k\ge 4$ the parafermionic part is not any longer a unitary model
of the Virasoro algebra and we do not know in general  the split 
of the parafermionic characters $\kappa^l_0$ in $\sigma^*$ even
and $\sigma^*$ odd parts.
The only other model where we succeeded in finding this decomposition
is the $k=6$ unitary model.

\subsec{Action of $\sigma^*$ for the (k=6) model}

For $k=6$ the central charge is $c={9\over 4}$ so that the parafermionic 
theory contributes $c={5\over 4}$. Even though this number is not
contained in the unitary series of the ${\cal N}=0$ Virasoro
algebra it is a member of the unitary series of the ${\cal N}=1$
super Virasoro algebra at $m=6$
\eqn\super{   c={3\over 2}\left( 1-{8\over m(m+2)} \right) .}
Some inspection reveals that the vacuum character of the
$k=6$ parafermionic theory can be written as
\eqn\vaccumc{
             \kappa^0_0={1\over 2}\left(\chi^{NS}_0+\chi^{\widetilde{NS}}_0
                   +\chi^{NS}_3+\chi^{\widetilde{NS}}_3  \right), }            
where again the lower indices denote the conformal dimensions
of the HWRs. 
Expanding the superconformal characters we find for the
first eight levels
\eqn\expand{\eqalign{
      {1\over 2}\left( \chi^{NS}_0+\chi^{\widetilde{NS}}_0\right)
  &=q^{-{5\over 96}}\left(1+q^2+q^3+3q^4+3q^5+7q^6+8q^7+15q^8+\ldots\right) \cr
     {1\over 2} \left( \chi^{NS}_3+\chi^{\widetilde{NS}}_3\right)
      &=q^{-{5\over 96}}\left(q^3+q^4+3q^5+4q^6+7q^7+10q^8+\ldots\right). \cr}}
Thinking of these states as being generated by $L_m$ and $G^\pm_r$
the number of states at each mass level in the first row of \expand\ 
agrees with the
expected number of $\sigma^*$ even states, whereas the 
number of states in the second row completely agrees
with the expected number of $\sigma^*$ odd states.
Thus we claim that this pattern will continue  to all mass levels and that
the decomposition of the parafermionic vacuum character in terms
of ${\cal N}=1$ super Virasoro characters automatically reflects
 the desired
split into $\sigma^*$ even and $\sigma^*$ odd states.
The remaining $\kappa^l_0$ characters have a similar decomposition
\eqn\expandb{\eqalign{
\kappa^2_0&={1\over 2}\left(\chi^{NS}_{1\over 4}+
                           \chi^{\widetilde{NS}}_{1\over 4}+
                   \chi^{NS}_{5\over 4}+
                  \chi^{\widetilde{NS}}_{5\over 4}  \right) \cr
 \kappa^4_0&={1\over 2}\left(\chi^{NS}_{1\over 4}-
                           \chi^{\widetilde{NS}}_{1\over 4}+
                   \chi^{NS}_{5\over 4}-
                  \chi^{\widetilde{NS}}_{5\over 4}  \right) \cr     
\kappa^6_0&={1\over 2}\left(\chi^{NS}_0-\chi^{\widetilde{NS}}_0+
                   \chi^{NS}_3-\chi^{\widetilde{NS}}_3  \right). \cr }  }    
Analogously to the vacuum character we conjecture that the action
of charge conjugation on these characters is simply
\eqn\expandc{\eqalign{
\chi^0_{0,0}(\sigma^*)&=\sqrt{ 2\eta\over \theta_2 }
                {1\over 2}\left[\left(\chi^{NS}_0-\chi^{NS}_3\right)
                   +\left(\chi^{\widetilde{NS}}_0
                   -\chi^{\widetilde{NS}}_3\right)  \right]       \cr
\chi^2_{0,0}(\sigma^*)&=\sqrt{ 2\eta\over \theta_2 }
          {1\over 2}\left[\left(\chi^{NS}_{1\over 4}-
                      \chi^{NS}_{5\over 4}\right)+
                        \left(   \chi^{\widetilde{NS}}_{1\over 4}-
                  \chi^{\widetilde{NS}}_{5\over 4}  \right)\right] \cr
\chi^4_{0,0}(\sigma^*)&=\sqrt{ 2\eta\over \theta_2 }
               {1\over 2}\left[\left(\chi^{NS}_{1\over 4}-
                      \chi^{NS}_{5\over 4}\right)-
                        \left(   \chi^{\widetilde{NS}}_{1\over 4}-
                  \chi^{\widetilde{NS}}_{5\over 4}  \right)\right] \cr
\chi^6_{0,0}(\sigma^*)&=\sqrt{ 2\eta\over \theta_2 }
                {1\over 2}\left[\left(\chi^{NS}_0-\chi^{NS}_3\right)
                   -\left(\chi^{\widetilde{NS}}_0
                   -\chi^{\widetilde{NS}}_3\right)  \right] .\cr }}
In section 6 we will see that this guess is supported by the
consistency of the results we will obtain for the $\ZZ_2$ twisted
sectors of the $(6)^4$ and the $(2)^3 (6)^2$ Gepner models.
The results derived in this section will provide the main technical
information we need in order to construct the complete partition
functions of the anti-holomorphic orbifold models.

\subsec{Speculations about the $\sigma^*$ action at arbitrary  level}

Unfortunately, we have not managed yet to find the decomposition
of the uncharged characters into $\sigma^*$ even and $\sigma^*$ odd
parts for generic level $k$. Apparently, this a pure CFT problem. 
In \runify\ the ${\cal W}$-algebra for the $\sigma^*$ even
part of the parafermionic CFT was determined to be a
${\cal W}(2,4,6,8,10)$ algebra. The structure constants of this ${\cal
W}$-algebra are not known explicitly. For
$c={4\over 5}$ we expect this ${\cal W}$ algebra to truncate
to the Virasoro algebra and for $c={5\over 4}$ we expect it
to truncate to the ${\cal W}(2,4,6)_3$ algebra, which is the 
bosonic projection of the ${\cal N}=1$ Virasoro algebra.

Furthermore, the ${\cal W}(2,4,6,8,10)$ algebra should admit
rational models for the unitary series of the parafermions
$c={2(k-1)\over k+2}$, where in addition for all these values of $c$ a HWR 
of conformal dimension $h=3$ should appear.
Then the parafermionic vacuum character would split like
\eqn\splitf{ \kappa^0_0=\chi_0 + \chi_3 }
and the action of $\sigma^*$ would simply be
\eqn\splitact{ \kappa^0_0(\sigma^*)=\chi_0 - \chi_3 .}
Classifying the minimal models of this ${\cal W}(2,4,6,8,10)$-algebra 
and derive their modular
transformation properties would be the main task   
on the way to the computational exploration of hundreds of different 
Gepner models.

\newsec{SCFT for the $(\IP_4[5]\times S^1)/\ZZ_2$ model}

Since we are equipped with the action of  $\sigma^*$ on all states
in the Hilbert space of the  
$k=3$ unitary model, we are in the position to compute the 
one-loop partition function for the Gepner type model defined by
\eqn\quin{    {(3)^5 \times S^1\over \sigma^*} ,}
which is expected to geometrically correspond to the $G_2$ manifold
\eqn\quinb{    {\IP_4[5] \times S^1\over \sigma^*} .}
Remember, that $\sigma^*$ acts  by charge conjugation on the 
${\cal N}=2$ unitary factor models and by inversion $y\to -y$ on the
coordinate compactified on a circle. 

In the following we will demonstrate the construction of modular
invariant partition functions for $G_2$ manifolds on this specific
examples. The generalization to other Gepner models is straightforward. 

As in section 3 we are  now considering the Gepner model defined 
by the tensor product of five copies of the $k=3$ unitary model and
two free fermions forming the $SO(2)_1$ current algebra. 

\subsec{The free fermion part}

Since we
now compactify one more direction on a circle, we first have to
decompose the $SO(2)_1$ representations into $SO(1)_1\times SO(1)_1$
representations, where by $SO(1)_1$ we mean the CFT of one free fermion,
i.e. the Ising model. The first $SO(1)_1$ factor corresponds to
the flat direction and the second $SO(1)_1$ factor to the direction
compactified on $S^1$. 
The characters of the Ising model are
\eqn\charrepb{\eqalign{
  O_1&={1\over 2}\left( \sqrt{\theta_3\over \eta} +
                         \sqrt{\theta_4\over \eta}\right) \cr
  V_1&={1\over 2}\left( \sqrt{\theta_3\over \eta} -
                  \sqrt{\theta_4\over \eta}\right) \cr
    S_1&=\sqrt{\theta_2\over 2\eta }. \cr }}
Starting with the $SO(1)_1\times SO(1)_1$ theory, the
characters of the $SO(2)_1$ CFT are given by taking orbits under
the simple current $J=(V_1\, V_1)$
\eqn\spliti{\eqalign{      
            O_2&=O_1\, O_1 + V_1\, V_1 \cr
            V_2&=O_1\, V_1 + V_1\, O_1 \cr
            S_2&=S_1\, S_1  \cr 
            C_2&=S_1\, S_1  .\cr }} 
Due to the fusion rule $S_1\times V_1=S_1$ the spinor representation 
$S_1$ with conformal
dimension $h={1\over 16}$ is a fixed point under the action of the
simple current and therefore  $(S_1\,S_1)$ gives rise to the representations
$S_2$ and $C_2$.
Under a modular $S:\tau\to -1/\tau$ transformation
the three representations $(O_1,V_1,S_1)$ of the free fermion CFT transform as
\eqn\smatric{
                 S={1\over 2}\left(\matrix{ 1 & 1 & \sqrt{2} \cr
                                            1 & 1 & -\sqrt{2} \cr 
                                         \sqrt 2 & -\sqrt 2 & 0 \cr }\right).}
Note that under the action of $\sigma^*$ the characters $O_1$ is invariant, 
whereas $V_1$ is mapped to $-V_1$. Additionally, $\sigma^*$ exchanges
the spinor $S_2$ representation with the anti-spinor $C_2$
representation, so that it also acts as charge conjugation
in the free fermion part of the CFT.

\subsec{Space-time supersymmetry}

We expect that under the action of $\sigma^*$ half of the supersymmetry
is broken, so that we are left with $N=2$ supersymmetry in three dimensions. 
Indeed the gravitinos from equation \gravi\ split into
\eqn\gravib{\eqalign{ &\left[(0,0,0)^5 \otimes (V_1 O_1+O_1 V_1) \right]_L
             \times \left[ (0,1,1)^5\otimes (C_2) +
                           (3,4,1)^5\otimes (S_2)\right]_R, \cr 
            &\left[ (0,1,1)^5\otimes (C_2) +
                   (3,4,1)^5\otimes (S_2)  \right]_L\times
            \left[ (0,0,0)^5 \otimes (V_1 O_1 + O_1 V_1) \right]_R }}
and are mapped under $\sigma^*$ to
\eqn\gravic{\eqalign{ &\left[ (0,0,0)^5 \otimes (-V_1 O_1 + O_1 V_1) \right]_L
             \times \left[ (3,4,1)^5\otimes (S_2) +
                        (0,1,1)^5\otimes (C_2) \right]_R, \cr 
            &\left[ (3,4,1)^5\otimes (S_2) +
                   (0,1,1)^5\otimes (C_2) \right]_L\times
             \left[ (0,0,0)^5 \otimes (-V_1 O_1 + O_1 V_1) \right]_R }}
so that precisely half of the supercharges are projected out. 
After the $\sigma^*$ projection the vacuum orbit contributes
the three dimensional ${\cal N}=2$ supergravity multiplet in 
addition to one chiral multiplet. 

Moreover, the states $[(3,3,0)^5\otimes (O_1\, O_1)]_{L,R}$ of conformal
dimension $(h,q)=(3/2,3)$ are mapped under $\sigma^*$
to their charge conjugate   states 
$[(3,-3,0)^5\otimes (O_1\, O_1)]_{L,R}$ so that the
symmetric linear combinations survive. Note  that these linear combinations
correspond to the covariantly constant three-form on the $G_2$ manifold. 

\subsec{Massless states in the $\ZZ_2$ untwisted sector}

The three dimensional massless spectrum appearing in the $\ZZ_2$ untwisted
sector is quite general. In fact the $\sigma^*$ action on massless
chiral or anti-chiral states exchanges
\eqn\chiral{ \left[ (h=1/2,q=1)\otimes (O_1 O_1 + V_1 V_1) \right]_L \times
                  \left[ (h=1/2,q=\pm1) \otimes (O_1 O_1 + V_1 V_1) \right]_R }
with its charge conjugate
\eqn\chiral{ \left[ (h=1/2,q=-1)\otimes (O_1 O_1 - V_1 V_1) \right]_L \times
                \left[ (h=1/2,q=\mp 1) \otimes (O_1 O_1 - V_1 V_1) \right]_R }
and therefore leads to one chiral multiplet for both the Type IIA
and the Type IIB string. 
Thus, from the untwisted sector we get the supergravity multiplet
and $h_{11}+h_{21}+1$ chiral multiplets.
This is in agreement with the result from the geometric 
computation where the number of chiral multiplets is given by
$b_2+b_3=h_{11}+h_{21}+1$.
Since the uncharged states are not touched at all, 
performing the mirror sign flip  $U(1)_R\to -U(1)_R$ in the $c=9$ Gepner model
does  lead to isomorphic SCFTs after the $\ZZ_2$ orbifold, as well.

\subsec{Partition function in the $\ZZ_2$ twisted sector}

The recipe for finding the partition function in the $\ZZ_2$ twisted
sector is to  first compute the trace with the $\sigma^*$ insertion
${\scriptstyle\sigma^*}\square\limits_1$ and then apply a modular 
S-transformation
to get the sector $1\square\limits_{\sigma^*}$\ . 
For the $1\square\limits_{\sigma^*}$ sector a number of non-trivial 
consistency conditions arise. First, this sector must be level matched
which means that all twisted states must satisfy  
\eqn\levelm{    h_L-h_R\in\ZZ/2 .}
Second, the twisted sector must really admit the interpretation
as a partition function with non-negative integer coefficient.  
Third,  the twisted sector must vanish and there must be the same
number of space-time bosons and fermions. Finally, there must not be any
tachyons in the model. 
The partition function in the $\ZZ_2$ twisted sector is then
\eqn\twparti{   {1\over 2}\left(  1\square\limits_{\sigma^*} +
                   {\scriptstyle\sigma^*}\square\limits_{\sigma^*} \right) }
where ${\scriptstyle\sigma^*}\square\limits_{\sigma^*}$ can be obtained from
$1\square\limits_{\sigma^*}$ by applying a modular $T:\tau\to\tau+1$
transformation. 

As we argued in section 3 only uncharged states, counted in 
the $\chi^l_{0,0}$ characters  for  each factor theory, can contribute
to the ${\scriptstyle\sigma^*}\square\limits_1$ sector. 
For $k=3$ there are only two
uncharged characters, $\chi^0_{0,0}$ and $\chi^2_{0,0}$, which appear
in different orbits under the GSO simple current \simple.
Note that all these uncharged states are combined with  the vector 
representation, $V_2$
of $SO(2)$. Moreover, only the trivial Kaluza-Klein and winding
mode for the compact $S^1$ direction is invariant under
$\sigma^*$, so that  
collecting everything together we obtain the following
partition function 
\eqn\insert{ 
   {\scriptstyle\sigma^*}\square\limits_1={2\over |\eta |^2}\, 
        \left|\sqrt{\eta\over \theta_2}\right|^2\, 
           \left| O_1 V_1 - V_1 O_1 \right|^2\,
           \sum_{n=0}^5  {\ts{5\choose n}} \left| 
           \left( \chi^0_{0,0}(\sigma^*) \right)^n\,
            \left( \chi^2_{0,0}(\sigma^*) \right)^{5-n} \right|^2 .}
For the free fermion part we are carefully treating  the order of the
two $SO(1)$ factors, whereas for simplicity for the five $k=3$ tensor factors
we introduced a combinatorial degeneracy. Using equation \sigmadrei\
yields
\eqn\insert{ 
   {\scriptstyle\sigma^*}\square\limits_1={2^6\over |\eta |^2}\, 
        \left|\sqrt{\eta\over \theta_2}\right|^{12}\, 
           \left| O_1 V_1 - V_1 O_1 \right|^2\,
           \sum_{n=0}^5  {\ts{5\choose n}} \left| 
           \left( \chi_0-\chi_3\right)^n\,
            \left( \chi_{2\over 5}-\chi_{7\over 5}\right)^{5-n} \right|^2 .}
The next step is to apply a modular S transformation. 
For the free fermions we obtain
\eqn\frees{ (O_1 V_1 - V_1 O_1) \to {1\over \sqrt 2}\bigl(
              S_1(O_1+V_1)-(O_1+V_1) S_1 \bigr) }
which apparently vanishes and also guarantees that the number of
space-time bosons contained in the first term, $ S_1(O_1+V_1)$, 
agrees with the number of space-time fermions counted 
in the second term $(O_1+V_1) S_1$. 
Using the formulae collected in appendix A, 
under a modular S-transformation the characters of the three states
Potts model behave as follows
\eqn\pottss{\eqalign{
        \left(\chi_0-\chi_3\right)&\to 
    {\ts\sqrt{4\over 5}}\left( \sin\left({\ts{\pi\over 5}}\right)\,
                           (\chi_{1\over 8}+\chi_{13\over 8})
                +  \sin\left({\ts{3\pi\over 5}}\right)\,
                           (\chi_{1\over 40}+\chi_{21\over 40}) \right)\cr
        \left(\chi_{2\over 5}-\chi_{7\over 5}\right)&\to 
       {\ts\sqrt{4\over 5}}\left( -\sin\left({\ts{3\pi\over 5}}\right)\,
                           (\chi_{1\over 8}+\chi_{13\over 8})
                +  \sin\left({\ts{\pi\over 5}}\right)\,
                           (\chi_{1\over 40}+\chi_{21\over 40}) \right).\cr }}
Thus, for the twisted sector partition function we get an expression
of the form
\eqn\twist{\eqalign{  1\square\limits_{\sigma^*}=
            {2^5\over |\eta |^2}\, 
        \left|\sqrt{\eta\over \theta_4}\right|^{12}\, 
           &\bigl| S_1(O_1+V_1)-(O_1+V_1) S_1 \bigr|^2\, \cr
   &\sum_{n=0}^5  {\ts{5\choose n}} \left| 
           \left( \chi_{1\over 40}+\chi_{21\over 40}\right)^n\,
        \left( \chi_{1\over 8}+\chi_{13\over 8}\right)^{5-n} \right|^2.\cr }}
Note that all terms in the sum satisfy the 
level matching condition
\eqn\lequui{   h_L-h_R\in {\ZZ\over 2}   ,} 
which we consider as a highly non-trivial check that our $\sigma^*$
action in the $k=3$ unitary model is indeed correct. Moreover, the  
computed twisted sector partition function really has the 
interpretation as a trace over states in a new sector of the Hilbert space.
All the requirements we mentioned at the beginning of this subsection
are therefore satisfied. 

Since the sum in \twist\ starts like
\eqn\summe{  \left| (\chi_{1\over 40}+\chi_{21\over 40})^5 \right|^2 
       + {\rm higher\ terms} }
the ground state energy in the $\sigma^*$ twisted sector
is
\eqn\ground{   E={1\over 16}   }
and there do not appear any new massless states.
Thus, the overall massless spectrum of the ${(3)^5 \times S^1\over \sigma^*}$
model is
\eqn\massl{ {\rm SUGRA}+103_u\ {\rm chiral\ multiplets} .}
We conclude that at the Gepner point of the quintic $\sigma^*$ acts freely, 
which has to be compared with what happens in the large radius limit 
where we can describe the model geometrically.

\newsec{Geometric interpretation of the $(\IP_4[5]\times S^1)/\ZZ_2$ model}

\subsec{The Geometry}

The antiholomorphic involution on the Calabi-Yau manifold is induced
from the ambient projective space. Since the different $k=3$ factors
are not interchanged by the orbifold the involution is the obvious
$z_i\to \bar{z}_i$ (the A--type involution in the language of
\rpioline{}). The Fermat quintic
\eqn\quintic{ \lbrace z_0^5 + z_1^5 + z_2^5 + z_3^5 + z_4^5 = 0\rbrace 
\in \IP_{1,1,1,1,1}} 
is obviously mapped to itself by the involution. 

Now we want to determine the de Rahm cohomology of the quotient, and
for this we have to find the invariant classes on the initial space
$\IP_4[5]\times S^1$. But the classes on a Cartesian product are
simply the product of the cohomology classes and it suffices to
discuss the forms on the two factors separately:

\def\litem{\par\noindent \hangindent=\parindent\ltextindent}
\def\ltextindent#1{\hbox to \hangindent{$\quad$#1\hss}\ignorespaces}

{\parindent=4cm
\litem{${\rm d}y \in H^1(S^1)$}
  The volume form on the $S^1$ is odd under $\sigma^\ast$.
\litem{$\omega \in H^{1,1}(\IP_4[5])$}
  The K\"ahler form is induced from the ambient space. Therefore it is
odd under the antiholomorphic involution since the K\"ahler form on
$\IP_4$ is odd. (In our examples this will always be the case for all
$h^{11}$ classes)
\litem{$\Omega, \bar{\Omega}\in H^{3,0}\oplus H^{0,3}$}
  The involution exchanges $H^{3,0}$ and $H^{0,3}$, so there is one
invariant and one antiinvariant combination. We choose the phase of
$\Omega$ such that $\Re(\Omega)$ is invariant.
\litem{$\eta_i \in H^{2,1}\oplus H^{1,2}$}
  By the same argument $h^{21}=101$ of the $2h^{21}$ forms are even, 
say $\eta_i$ with $i=1,\dots,h^{21}$.
\par}
\noindent The nontrivial Betti numbers of the quotient $(\IP_4[5]\times
S^1)/\sigma^\ast$ are then
{\parindent=4cm
\litem{$b^1=0$}
  since ${\rm d}y$ is projected out. This is necessary for $G_2$
holonomy. 
\litem{$b^2=0$}
  since there are no invariant $2$--forms.
\litem{$b^3=103$}
  from the $101$ invariant forms $\eta_i$, the form ${\rm d}y
\wedge \omega$ and from $\Re(\Omega)$.
\par}

\subsec{Singularities}

Away from the fixed points of the $\sigma^\ast$ action the quotient is
a manifold. But the geometric group action has fixed loci that give
rise to $A_1$ singularities. 
The fixed points on $\IP_4[5]\times S^1$ are the product of the
individual fixed point loci. Let $\Sigma \in \IP_4[5]$ be the real
points of the quintic, then the overall fixed set is $(\Sigma \times
\lbrace 0 \rbrace) \cup (\Sigma \times \lbrace {1 \over
2}\rbrace)$. 

$\Sigma$ is the solution of the real polynomial equation
$x_1^5+\cdots+x_5^5=0$ in $\IRP_4$. We want to determine its
topology: First note that $\IR\to\IR, x\mapsto x^5$ is bijective, so
we might just as well analyze $x_1+\cdots+x_5=0$ in $\IRP_4$. On the
double cover $S^4$ of $\IRP_4$ this equation determines an equatorial
$S^3$ (the intersection of one $4$--plane with the unit $S^4$ embedded
in $\IR^5$). Finally, modding  out the remaining (antipodal) $\ZZ_2$--action
leads to $\Sigma \simeq \IRP_3$.
Locally the involution acts as $(\xi_1, \xi_2, \xi_3, y) \to
(\bar{\xi}_1, \bar{\xi}_2, \bar{\xi}_3, -y)$. So it leaves $\Re \xi_1,
\Re \xi_2, \Re \xi_3$ invariant and inverts the 4 other directions. So
the normal direction over the fixed set looks like $\IR^4/\ZZ_2$.

We have characterized the singularities as two disjoint copies of
$\IRP_3$ with $A_1$ normal bundle. Can one resolve these
singularities? It is anticipated by \refs{\rjoyce\rkehagiasa} that this is
not possible within $G_2$ holonomy, but no proof is available. However
the following argument shows that the usual resolution (gluing
Eguchi--Hansen spaces for the $A_1$--singularities) is not possible.

The Eguchi--Hansen space contains a nontrivial $S^2$ whose size must
become a modulus of the resulting $G_2$ manifold. But the moduli space
of $G_2$ metrics is $b_3$--dimensional, so the resolution must
increase $b_3$. This can only happen if the singular locus has $b_1>0$
because the $S^2$ has to combine with a $1$--cycle in the base to form
a $3$--cycle. But the $\IRP_3$ has $b_1=0$. One might hope that a flop
(see \ramv)
might ameliorate the situation but this will not help in the present
case: it would only exchange $\IRP_3\times \IR^4/\ZZ_2
\longleftrightarrow \IR^4/\ZZ_2 \times \IRP_3$

Nevertheless, applying an adiabatic argument we expect precisely
one massless chiral multiplet from the 
local geometry around the
singularity. This multiplet arises from dimensionally reducing 
the six-dimensional abelian gauge field
living at the orbifold fixed point of $\IR^4/\ZZ_2$ 
on $\IRP_3$. 
Thus, there is a clear mismatch
between the CFT and the geometric computation. 
The other examples studied in the next section feature a similar
discrepancy of the twisted sector massless modes. 
We will resolve this puzzle after we present two more SCFT examples.

\newsec{SCFTs for the $(\IP_{1,1,1,1,4}[8]\times S^1)/\ZZ_2$ and
        $(\IP_{1,1,2,2,2}[8]\times S^1)/\ZZ_2$ model  }

Since, we  know the action of $\sigma^*$ for the $k=6$ 
unitary model, the two models
\eqn\quinr{    {(6)^4 \times S^1\over \sigma^*}, \quad\quad
               {(2)^3\, (6)^2 \times S^1\over \sigma^*} }
are also calculable. Geometrically, these two Gepner models
correspond to the $G_2$ manifolds
\eqn\quinbv{    {\IP_{1,1,1,1,4}[8]_{1,149} \times S^1\over \sigma^*},
             \quad\quad
                {\IP_{1,1,2,2,2}[8]_{2,86} \times S^1\over \sigma^*} ,}
where the index denotes the Hodge numbers $(h_{11},h_{21})$. 

\subsec{ $(\IP_{1,1,1,1,4}[8] \times S^1)/\ZZ_2$ }

In complete analogy to the quintic the massless spectrum in 
the $\ZZ_2$ untwisted sector consists of the ${\cal N}=2$ supergravity
multiplet in addition to 151 chiral multiplets. 
In the ${\scriptstyle{\sigma^*}}\square\limits_1$ sector 
the four uncharged characters
$\chi^{2j}_{0,0}$ for $j\in\{0,1,2,3\}$ can appear.
In contrast to the $k=3$ Gepner model here the pairs of characters 
$(\chi^0_{0,0},\chi^6_{0,0})$  and $(\chi^2_{0,0},\chi^4_{0,0})$ appear in the
same orbit under the spectral flow. Therefore the 
${\scriptstyle{\sigma^*}}\square\limits_1$
partition function reads
\eqn\sechsa{\eqalign{ 
   {\scriptstyle\sigma^*}\square\limits_1={2\over |\eta |^2}\, 
        \left|\sqrt{\eta\over \theta_2}\right|^{2}\, 
           \bigl| O_1 V_1 - V_1 O_1 \bigr|^2\,
           {1\over 2} \sum_{i,j,k,l=0}^3  \Bigl|& 
           \chi^{2i}(\sigma^*) \chi^{2j}(\sigma^*) \chi^{2k}(\sigma^*) 
           \chi^{2l}(\sigma^*)+ \cr
    &\chi^{6-2i}(\sigma^*) \chi^{6-2j}(\sigma^*) 
         \chi^{6-2k}(\sigma^*) \chi^{6-2l}(\sigma^*) \Bigr|^2. \cr }}
Inserting \expandc\ and using the modular transformation properties summarized
in appendix B 
\eqn\susyms{\eqalign{
        (\chi^{NS}_0-\chi^{NS}_3)&\to 
           {\ts{1\over \sqrt 2}}(\chi^{NS}_{1\over 32}+\chi^{NS}_{33\over 32})
                           + \chi^{NS}_{5\over 32}  \cr
     (\chi^{NS}_{1\over 4}-\chi^{NS}_{5\over 4})&\to 
           {\ts{1\over \sqrt 2}}(\chi^{NS}_{1\over 32}+\chi^{NS}_{33\over 32})
                           - \chi^{NS}_{5\over 32}  \cr
      (\chi^{\widetilde{NS}}_0-\chi^{\widetilde{NS}}_3)&\to 
                 c_1 (\chi^R_{5\over 16}+\chi^R_{29\over 16})
                           + c_2 (\chi^R_{1\over 16} + \chi^R_{9\over 16})\cr
     (\chi^{\widetilde{NS}}_{1\over 4}-\chi^{\widetilde{NS}}_{5\over 4})&\to 
                 -c_2 (\chi^R_{5\over 16}+\chi^R_{29\over 16})
                           + c_1 (\chi^R_{1\over 16} + \chi^R_{9\over 16})\cr}}
with 
\eqn\coeff{ c_1={\ts{1\over \sqrt{6}}}
     \left(\cos\left({\ts{5\pi\over 24}}\right)-
         \cos\left({\ts{11\pi\over 24}}\right) \right),
                  \quad\quad
        c_2={\ts{1\over \sqrt{6}}}\left(\cos\left({\ts{\pi\over 24}}\right)+
                     \cos\left({\ts{7\pi\over 24}}\right) \right).}
we obtain for the twisted sector partition function
\eqn\sechsb{ 
   1\square\limits_{\sigma^*}={2^4\over |\eta |^2}\, 
        \left|\sqrt{\eta\over \theta_4}\right|^{10}\, 
           \bigl| S_1(O_1+V_1)-(O_1+V_1) S_1  \bigr|^2\,
          I(q,\overline{q}).  }
with
\eqn\sechsf{\eqalign{ I(q,\overline{q})=& 
{\ts{1\over 128}} \left| (\chi^R_{1\over 16} + \chi^R_{9\over 16})^4\right|^2+ 
{\ts{1\over 8}} \left| (\chi^{NS}_{1\over 32}+\chi^{NS}_{33\over 32})^4 
\right|^2 +
{\ts{3\over 16}} \left| (\chi^R_{1\over 16} + \chi^R_{9\over 16})^2
                (\chi^{NS}_{1\over 32}+\chi^{NS}_{33\over 32})^2\right|^2+\cr
&{\ts{1\over 32}} \left| (\chi^R_{1\over 16} + \chi^R_{9\over 16})^3
                   (\chi^R_{5\over 16}+\chi^R_{29\over 16}) \right|^2+
{\ts{3\over 8}} \left| (\chi^R_{1\over 16} + \chi^R_{9\over 16})
                  (\chi^{NS}_{1\over 32}+\chi^{NS}_{33\over 32})^2
                  (\chi^R_{5\over 16}+\chi^R_{29\over 16})  \right|^2+\cr
&{\ts{3\over 64}} \left| (\chi^R_{1\over 16} + \chi^R_{9\over 16})^2
                   (\chi^R_{5\over 16}+\chi^R_{29\over 16})^2  \right|^2+ 
{\ts{3\over 16}} \left| (\chi^{NS}_{1\over 32}+\chi^{NS}_{33\over 32})^2
                   (\chi^R_{5\over 16}+\chi^R_{29\over 16})^2  \right|^2+ \cr
&{\ts{1\over 32}} \left| (\chi^R_{1\over 16} + \chi^R_{9\over 16})
                   (\chi^R_{5\over 16}+\chi^R_{29\over 16})^3 \right|^2+
{\ts{1\over 128}} \left| (\chi^R_{5\over 16} + \chi^R_{29\over 16})^4
 \right|^2 + \cr
&{\ts{3\over 4}} \left| (\chi^R_{1\over 16} + \chi^R_{9\over 16})^2 
                  (\chi^{NS}_{1\over 32}+\chi^{NS}_{33\over 32})
                  (\chi^{NS}_{5\over 32}) \right|^2 +
 \left| (\chi^{NS}_{1\over 32}+\chi^{NS}_{33\over 32})^3
                  (\chi^{NS}_{5\over 32}) \right|^2 + \cr
&{\ts{3\over 2}} \left| (\chi^R_{1\over 16} + \chi^R_{9\over 16})      
                  (\chi^{NS}_{1\over 32}+\chi^{NS}_{33\over 32})
                  (\chi^R_{5\over 16}+\chi^R_{29\over 16})
                  (\chi^{NS}_{5\over 32}) \right|^2 +
{\ts{3\over 4}} \left| (\chi^R_{1\over 16} + \chi^R_{9\over 16})^2
                  (\chi^{NS}_{5\over 32})^2 \right|^2 \cr
&{\ts{3\over 4}} \left| (\chi^{NS}_{1\over 32}+\chi^{NS}_{33\over 32})
                  (\chi^R_{5\over 16}+\chi^R_{29\over 16})^2
                  (\chi^{NS}_{5\over 32}) \right|^2 +
{\ts{3}} \left| (\chi^{NS}_{1\over 32}+\chi^{NS}_{33\over 32})^2
           (\chi^{NS}_{5\over 32})^2 \right|^2 +\cr
&{\ts{3\over 2}} \left| (\chi^R_{1\over 16} + \chi^R_{9\over 16})
                  (\chi^R_{5\over 16}+\chi^R_{29\over 16})
                  (\chi^{NS}_{5\over 32})^2 \right|^2 +
{\ts{3\over 4}} \left| (\chi^R_{5\over 16}+\chi^R_{29\over 16})^2
                  (\chi^{NS}_{5\over 32})^2 \right|^2 +\cr
&{\ts{4}} \left| (\chi^{NS}_{1\over 32}+\chi^{NS}_{33\over 32})
           (\chi^{NS}_{5\over 32})^3 \right|^2+
{\ts{2}} \left| (\chi^{NS}_{5\over 32})^4\right|^2. \cr }}
Again, the expression satisfies level matching, Bose-Fermi
degeneracy and absence of tachyons. 
Since all coefficient in \sechsf\ are of the form
\eqn\aform{   a_i={N\over 128}, \quad {\rm with}\ N\in\ZZ_+ }
and taking into account that the Ramond ground states in 
$\chi^R_{1\over 16}$, $\chi^R_{9\over 16}$,
$\chi^R_{5\over 16}$ and $\chi^R_{29\over 16}$ are twofold degenerate
the coefficient $2^4$ in \sechsb\ guarantees that the twisted sector
partition function  does 
indeed allow the interpretation as a trace over states in a Hilbert space.
The lowest energy states in  \sechsf\ are 
\eqn\loweste{  I(q,\overline q)=
 {1\over 8} \left|(\chi^{NS}_{1\over 32}+\chi^{NS}_{33\over 32})^4 \right|^2 + 
      {\rm higher\ terms} }
leading to a massless twisted sector ground state. 
This gives rise to {\bf 2} additional chiral multiplets in the
$\ZZ_2$ twisted sector.  
Thus, the overall massless spectrum of the ${(6)^4 \times S^1\over \sigma^*}$
model is
\eqn\massl{ {\rm SUGRA}+(151_u+2_t)\ {\rm chiral\ multiplets} .}
In contrast to the $(3)^5$ Gepner model here the $\sigma^*$ action has
 fixed points which give rise to new chiral multiplets in the 
twisted sector.

\subsec{ $(\IP_{1,1,2,2,2}[8] \times S^1)/\ZZ_2$ }

The massless spectrum in 
the $\ZZ_2$ untwisted sector consists of the ${\cal N}=2$ supergravity
multiplet in addition to 89 chiral multiplets.
Taking the uncharged characters of the $k=2$ and $k=6$ unitary models
and their behavior under the spectral flow into account we
arrive at the following expression for the 
${\scriptstyle{\sigma^*}}\square\limits_1$ partition function 
\eqn\hurraa{\eqalign{ 
   {\scriptstyle\sigma^*}\square\limits_1={2\over |\eta |^2}\, 
        \left|\sqrt{\eta\over \theta_2}\right|^{2}\, 
           \bigl| O_1 V_1 - V_1 O_1 \bigr|^2\,
           {1\over 2} \sum_{i,j,k=0}^1  \sum_{l,m=0}^3 \Bigl|& 
           \psi^{2i}(\sigma^*) \psi^{2j}(\sigma^*) \psi^{2k}(\sigma^*)\Bigl( 
           \chi^{2l}(\sigma^*)\chi^{2m}(\sigma^*)+ \cr
    &\chi^{6-2l}(\sigma^*) \chi^{6-2m}(\sigma^*) \Bigr) \Bigr|^2. \cr }}
We have denoted the characters of the $k=2$ unitary model by $\psi^l$ and
the characters of the $k=6$ unitary model by $\chi^l$.
After a modular S-transformation we obtain
for the twisted sector partition function
\eqn\hurrab{ 
   1\square\limits_{\sigma^*}={2^5\over |\eta |^2}\, 
        \left|\sqrt{\eta\over \theta_4}\right|^{12}\, 
           \bigl| S_1(O_1+V_1)-(O_1+V_1) S_1  \bigr|^2\,
          I(q,\overline q) .  }
with
\eqn\sechsf{\eqalign{ I(q,\overline{q})=& 
{\ts{1\over 64}} \left| (\psi_{0}+\psi_{1\over 2})^3
                   (\chi^R_{1\over 16} + \chi^R_{9\over 16})^2 \right|^2 +
{\ts{1\over 16}} \left| (\psi_{0}+\psi_{1\over 2})^3
                 (\chi^{NS}_{1\over 32}+\chi^{NS}_{33\over 32})^2 \right|^2+\cr
&{\ts{1\over 32}} \left| (\psi_{0}+\psi_{1\over 2})^3
                   (\chi^R_{1\over 16} + \chi^R_{9\over 16})
                   (\chi^R_{5\over 16}+\chi^R_{29\over 16}) \right|^2+
{\ts{1\over 64}} \left| (\psi_{0}+\psi_{1\over 2})^3
                   (\chi^R_{5\over 16}+\chi^R_{29\over 16})^2  \right|^2+\cr
&{\ts{1\over 4}} \left| (\psi_{0}+\psi_{1\over 2})^3
                  (\chi^{NS}_{1\over 32}+\chi^{NS}_{33\over 32})
                  (\chi^{NS}_{5\over 32})  \right|^2+
{\ts{1\over 4}} \left| (\psi_{0}+\psi_{1\over 2})^3
                  (\chi^{NS}_{5\over 32})^2  \right|^2+\cr
&{\ts{3\over 32}} \left| (\psi_{0}+\psi_{1\over 2})^2 (\psi_{1\over 16})
                   (\chi^R_{1\over 16} + \chi^R_{9\over 16})^2  \right|^2+
{\ts{3\over 8}} \left| (\psi_{0}+\psi_{1\over 2})^2 (\psi_{1\over 16})
                  (\chi^{NS}_{1\over 32}+\chi^{NS}_{33\over 32})^2  
 \right|^2+\cr
&{\ts{3\over 16}} \left| (\psi_{0}+\psi_{1\over 2})^2 (\psi_{1\over 16})
                   (\chi^R_{1\over 16} + \chi^R_{9\over 16})
                 (\chi^R_{5\over 16}+\chi^R_{29\over 16}) \right|^2+           
{\ts{3\over 32}} \left| (\psi_{0}+\psi_{1\over 2})^2 (\psi_{1\over 16})
                   (\chi^R_{5\over 16}+\chi^R_{29\over 16})^2 \right|^2+ \cr
&{\ts{3\over 2}} \left| (\psi_{0}+\psi_{1\over 2})^2 (\psi_{1\over 16})
                  (\chi^{NS}_{1\over 32}+\chi^{NS}_{33\over 32})
                  (\chi^{NS}_{5\over 32})  \right|^2+
{\ts{3\over 2}} \left| (\psi_{0}+\psi_{1\over 2})^2 (\psi_{1\over 16})
                  (\chi^{NS}_{5\over 32})^2  \right|^2+\cr
&{\ts{3\over 16}} \left| (\psi_{0}+\psi_{1\over 2}) (\psi_{1\over 16})^2
                   (\chi^R_{1\over 16} + \chi^R_{9\over 16})^2  \right|^2+
{\ts{3\over 4}} \left| (\psi_{0}+\psi_{1\over 2}) (\psi_{1\over 16})^2
                (\chi^{NS}_{1\over 32}+\chi^{NS}_{33\over 32})^2\right|^2+\cr
&{\ts{3\over 8}} \left| (\psi_{0}+\psi_{1\over 2}) (\psi_{1\over 16})^2
                   (\chi^R_{1\over 16} + \chi^R_{9\over 16})
                   (\chi^R_{5\over 16}+\chi^R_{29\over 16}) \right|^2+ 
{\ts{3\over 16}} \left| (\psi_{0}+\psi_{1\over 2}) (\psi_{1\over 16})^2
                   (\chi^R_{5\over 16}+\chi^R_{29\over 16})^2 \right|^2+ \cr
&{\ts{3}} \left| (\psi_{0}+\psi_{1\over 2}) (\psi_{1\over 16})^2
                  (\chi^{NS}_{1\over 32}+\chi^{NS}_{33\over 32})
                  (\chi^{NS}_{5\over 32})  \right|^2+
{\ts{3}} \left| (\psi_{0}+\psi_{1\over 2}) (\psi_{1\over 16})^2
                  (\chi^{NS}_{5\over 32})^2  \right|^2+\cr
&{\ts{1\over 8}} \left| (\psi_{1\over 16})^3
                   (\chi^R_{1\over 16} + \chi^R_{9\over 16})^2 \right|^2 +
{\ts{1\over 2}} \left| (\psi_{1\over 16})^3
               (\chi^{NS}_{1\over 32}+\chi^{NS}_{33\over 32})^2 \right|^2+\cr
&{\ts{1\over 4}} \left| (\psi_{1\over 16})^3
                   (\chi^R_{1\over 16} + \chi^R_{9\over 16})
                   (\chi^R_{5\over 16}+\chi^R_{29\over 16}) \right|^2+
{\ts{1\over 8}} \left| (\psi_{1\over 16})^3
                   (\chi^R_{5\over 16}+\chi^R_{29\over 16})^2  \right|^2+\cr
&{\ts{2}} \left| (\psi_{1\over 16})^3
                  (\chi^{NS}_{1\over 32}+\chi^{NS}_{33\over 32})
                  (\chi^{NS}_{5\over 32})  \right|^2+
{\ts{2}} \left| (\psi_{1\over 16})^3
                  (\chi^{NS}_{5\over 32})^2  \right|^2. \cr }}
Again, one can check that this expression satisfies 
level matching, Bose-Fermi degeneracy, absence of tachyons and the 
allowance of an interpretation as a trace. 
The contribution from the internal sector starts like  
\eqn\loweste{ I(q,\overline q)=
               {1\over 16} \left|\left(\psi_0+\psi_{1\over 2}\right)^3\, 
      \left(\chi^{NS}_{1\over 32}\right)^2 \right|^2 
                + {\rm higher\ terms} }
and gives rise to {\bf 2} additional  massless chiral multiplets in the
$\ZZ_2$ twisted sector. Summarizing, the massless spectrum of the 
${(2)^3(6)^2 \times S^1\over \sigma^*}$
model is
\eqn\masslb{ {\rm SUGRA}+(89_u+2_t)\ {\rm chiral\ multiplets} .}

Finally, we would like to mention that we have also computed the 
$(1)^9$ and $(2)^6$ Gepner models. These models do not have 
a geometric phase, as formally $h_{11}=0$. For the $(1)^9$ model
we found {\bf 85} chiral multiplets in the untwisted sector
and no massless states in the $\ZZ_2$ twisted sector. For the $(2)^6$ model
we found {\bf 91} chiral multiplets in the untwisted sector
and {\bf 2} additional massless states in the $\ZZ_2$ twisted sector.

\newsec{Geometric interpretation}

\subsec{The Geometry}

The discussion is very similar to the case with the quintic. One new
feature is the singular surface in the $\IP_{11222}$ which must be
blown up (One line in the moduli space of this model was investigated
in \rkehagiasa). Since the involution on the singular space extends to
an involution on the resolved space this poses no problem.  As in the
previous case the Betti numbers of the quotient are 
\eqn\bettif{b_1=0 \qquad b_2=0 \qquad b_3 = h^{11}+h^{21}+1}
The big difference is that the involution is free, that is it acts
without fixed points. The reason is that the Fermat polynomials have
only even degrees, and therefore no real solutions. For example the
fixed point set in $\IP_{11114}[8]$ is 
\eqn\emptycycle{\emptyset = 
  \lbrace x_1^8 + x_2^8 + x_3^8 + x_4^8 + x_5^2 = 0 \rbrace 
  \,\in\, \IRP_{11114}
}
Especially there should be no massless states in the twisted sector,
unlike in the SCFT analysis. 

In the remainder of this section we will discuss the topology in
greater detail with the ultimate goal to show that
$H_2(X;\ZZ)\not=\emptyset$. Unfortunately we will encounter technical
difficulties and our calculation works only for $h^{11}>1$, but we
expect it to hold in general.  This result is crucial for the physical
explanation that we will offer in the next section.

\subsec{Fundamental Group and Holonomy}

The following discussion is valid for all $X=(Y\times
S^1)/\ZZ_2$ that satisfy the following two conditions:
\item{(1)}$Y$ is a simply connected Calabi--Yau manifold
(e.g. a toric hypersurface) with Hodge numbers $h^{11}$, $h^{21}$.
\item{(2)}The $\ZZ_2$--action is free and inverts all classes in $H^{1,1}(Y)$.
\vskip0.5cm

The long exact homotopy sequence of the $\ZZ_2$--bundle 
$Y\times S^1\rightarrow X$
\eqn\leshtopy{
  \cdots \to 
  \underbrace{\pi_1(\ZZ_{2})}_{=0} \to 
  \underbrace{\pi_1(Y\times S^1)}_{=\ZZ} \to 
  \pi_1(X) \to 
  \underbrace{\pi_0(\ZZ_{2})}_{=\ZZ_{2}} \to
  \underbrace{\pi_0(Y\times S^1)}_{=0} \to 
  \cdots
}
implies that $\pi_1(X)$ must be infinite (Indeed 
$\pi_1(X)= \ZZ_2\ast\ZZ_2$).
But remember that for any compact manifold with torsion free
$G_2$--structure (i.e. the closed and coclosed $3$--form) the
following is equivalent:
$|\pi_1|<\infty \Leftrightarrow {\rm Hol}=G_2$. Since nevertheless
$b_1=0$ we conclude that
\eqn\hol{SU(3) \varsubsetneq {\rm Hol} \varsubsetneq G_2.}

\subsec{The Cartan--Leray spectral sequence}

One of the most useful tools to compute the homology of free quotients
of arbitrary spaces is the Cartan--Leray spectral sequence (see
\rkdbranes{} for a similar application). However it
turns out that there is an ambiguity which we cannot resolve. 
The result will be that $H_2(X;\ZZ)\not=0$ if $h^{11}>1$, and
undetermined otherwise. 

A spectral sequence is a systematic scheme to compute (co)homology
groups from other (hopefully more accessible) data. This particular
one is valid for arbitrary spaces (we choose $Y\times S^1$) with free
(proper) $\ZZ_2$--action (of course it is valid for more general
groups but this suffices for our purposes). It starts at
$E^2_{p,q}=\hat{H}_p(\ZZ_2,H_q(Y\times S^1,\ZZ))$ and converges to
$H_{p+q}(X;\ZZ)$, 
which is exactly what we are interested in.

In this sequence only non--negative $p$, $q$ have $E^2_{p,q}\not=0$ so
we can draw the initial data in the first quadrant. We do not assume
that the reader is familiar with group cohomology
$\hat{H}_p(\ZZ_2,F)$ --- $F$ denotes some $\ZZ_2$--module, that is
comes with an $\ZZ_2$--action like in our case the homology groups
$H_q(Y\times S^1,\ZZ)$. So let us quote the fundamental results:
{
\item{$\bullet$}$\hat{H}_p(\ZZ_2,\ZZ) = H_p(\IRP_\infty)$\quad
  (where $\ZZ$ has the trivial $\ZZ_2$--action)
\item{$\bullet$}
  $\hat{H}_p(\ZZ_2,\widetilde{\ZZ}) = 
  H_p(\IRP_\infty; \widetilde{\ZZ}) = 
  \big\{\ZZ_2, 0, \ZZ_2, 0, \dots$\hfill\break
  \hskip 1cm 
  where $\widetilde{\ZZ}$ are the integers with the nontrivial 
  $\ZZ_2$--action
\item{$\bullet$}$\hat{H}_p(\ZZ_2,0)=0$
\item{$\bullet$}$\hat{H}_0(\ZZ_2,F)=F\Big/\big<x-gx\big>$\quad
where $gx$ denotes the generator of $\ZZ_2$ acting on $x\in F$.
\par}
\vskip0.2cm
\noindent
For example we have (by assumption)
\eqn\coinvariantquotient{\eqalign{
gx=-x\quad&\forall x\in H_2(Y\times S^1;\ZZ)
\cr
\Rightarrow &\quad E^2_{0,2}=
\hat{H}_0\big(\ZZ_2, H_2(Y\times S^1;\ZZ)\big) = (\ZZ/2 \ZZ)^{h^{11}} = 
\ZZ_2^{h^{11}}
}}
So far we can evaluate 
\eqn\clEY{
E^2_{p,q}(X)= \qquad
\matrix{ 
\uparrow 
 &\strut\vrule& \ZZ_2^{h^{11}} &     0  & \ZZ_2^{h^{11}} & 0 \cr
q&\strut\vrule& \ZZ_2          &     0  & \ZZ_2          & 0 \cr
 &\strut\vrule& \ZZ            & \ZZ_2  & 0 & \ZZ_2\cr
\noalign{\hrule}
 &\strut\vrule& & & p & \rightarrow \cr
}}
Now starts the real work: At each entry in $E^2_{p,q}$ there is a map
that goes up $2-1$ and $2$ to the left. The cohomology at each point
is then $E^3_{p,q}$ and there are more maps, this time $3-1$ up and
$3$ left. This continues (``converges'') 
to $E^\infty_{p,q}$, whose $p+q={\rm const}$ 
diagonals are then the ``associated graded complex'' for
$H_{p+q}(X;\ZZ)$. The diagonal is empty if and only if the cohomology
$H_{p+q}(X;\ZZ)$ vanishes.

The only things that possibly influence the $p+q=2$ diagonal 
will be the maps
$d_2:E^2_{2,1}\to E^2_{0,2}$ and $d_3:E^3_{3,0}\to E^3_{0,2}$. 
Depending on the first map either
$E^3_{p,q}=\ZZ_2^{h^{11}}/\,{\rm img}\,d_2 = \ZZ_2^{h^{11}}$ 
or $\ZZ_2^{h^{11}-1}$. The second map $d_3$ can in principle 
also kill another $\ZZ_2$. 
Since we must assume the worst case this argument
only shows: If $h^{11}>2$ then $H_2(X;\ZZ)\not=\emptyset$.

\subsec{The $h^{11}=2$ case}

We can strengthen the result a little bit by further investigating the
topology of $X$. Via the projection on the first factor in $Y\times
S^1$ we see that $X$ is a $S^1$--bundle over $Y/\ZZ_2$. The topology
of the bundle is fixed by the fact that $X$ is orientable and
$Y/\ZZ_2$ is not. So we can also determine the (co)homology of $X$ by
first calculating the homology of $Y/\ZZ_2$ with the Cartan--Leray
spectral sequence and then use Leray's Theorem to compute the
cohomology of the bundle from the cohomology of the base and the
fiber.

The Cartan--Leray spectral sequence for $Y/\ZZ_2$ starts with
\eqn\clEX{
E^2_{p,q}(Y/\ZZ_2)= \qquad
\matrix{ 
\uparrow 
 &\strut\vrule& \ZZ_2^{h^{11}} &     0  & \ZZ_2^{h^{11}} & 0 \cr
q&\strut\vrule&    0           &     0  &     0          & 0 \cr
 &\strut\vrule& \ZZ            & \ZZ_2  & 0 & \ZZ_2\cr
\noalign{\hrule}
 &\strut\vrule& & & p & \rightarrow \cr
}}

Now the ambiguity is reduced to 
$H_2\left(Y/\ZZ_2;\ZZ\right) = \ZZ_2^{h^{11}} ~{\rm or}~
\ZZ_2^{h^{11}-1}$. In fact the only consistent possibilities are
\eqn\cohomologyYZ{
H_p(Y/\ZZ_2;\ZZ) = \left\{~
  \matrix{ 
    \strut                0 &                &\quad& p=6 \cr
    \strut            \ZZ_2 &                && p=5 \cr
    \strut     \ZZ^{h^{11}} &                && p=4 \cr
    \strut   \ZZ^{h^{21}+1} & [\oplus \ZZ_2] && p=3 \cr
    \strut \ZZ_2^{h^{11}-1} & [\oplus \ZZ_2] && p=2 \cr
    \strut            \ZZ_2 &                && p=1 \cr
    \strut              \ZZ &                && p=0 \cr
  }
\right.
}

Then Leray's Theorem amounts to a spectral sequence (this time for
cohomology) with
$E_2^{p,q}=H^p\big(Y/\ZZ_2; {\cal H}^q(S^1;\ZZ) \big)$. We find
\eqn\leray{
E_2^{p,q}(X)= \quad
\matrix{ 
\uparrow q
 &\strut\vrule& 
0  &
\ZZ_2 &
\ZZ^{h^{11}} &
\ZZ^{h^{21}+1} [\oplus \ZZ_2]  &
\ZZ_2^{h^{11}-1} [\oplus \ZZ_2]  &
\ZZ_2  &
\ZZ \cr
 &\strut\vrule& 
\ZZ &
0  &
\ZZ_2  &
\ZZ^{h^{21}+1} \oplus \ZZ_2^{h^{11}-1} [\oplus \ZZ_2]  &
\ZZ^{h^{11}}[\oplus \ZZ_2]  &
0  &
\ZZ_2  \cr
\noalign{\hrule}
 &\strut\vrule& & & & p  \rightarrow \cr
}}
The only interesting differential is $d_2:E_2^{1,1}\rightarrow
E_2^{3,0}$ and it has to vanish because the $\ZZ_2$ at $(p,q)=(1,1)$
has to survive as $H^2(X;\ZZ)=\ZZ_2\oplus \ZZ_2$. But then everything
survives to $E_\infty$ and we get
\eqn\lerayresult{
  H^3(X;\ZZ)_{\rm tor} = 
  H_2(X;\ZZ)_{\rm tor} = 
  \ZZ_2^{h^{11}-1} \quad {\rm or} \quad
  \ZZ_2^{h^{11}}
}

\newsec{Resolution of the Puzzle}

We have seen that the SCFT and the geometric analysis never yield the same
result. The resolution to this puzzle is that, as derived from mirror
symmetry \refs{\rcandelas,\raspinwall}, the Gepner model corresponds
to a point in the complexified K\"ahler moduli space where a non-trivial
background NS-NS two form flux, $B$,  has been turned on.
For instance, for the quintic Calabi-Yau it was shown that
the complex K\"ahler parameter of the Gepner model
is
\eqn\kahler{  B+iJ={1\over 2}+{i\over 2}\cot\left({\pi\over 5}\right) .}
Under the action of $\sigma^*$ the 2-cycle which is Poincare dual to the 
K\"ahler
form, $J$, combines with the 1-cycle of the additional $S^1$ to a 3-cycle
and gives rise to one chiral multiplet in three dimensions. 
The two-form, $B$, however is mapped to $-B$ under the action
of $\sigma^*$ and is therefore projected out. Thus, the two-form
is not any longer a continuous parameter. Due to the
fact that $B$ is only defined modulo one, there actually exist two
allowed discrete values for the background $B$ field, $B=0,1/2$.
This is very similar to what happens for compact Type I models
where also the NS-NS two-form does not survive the 
orientifold projection but nevertheless gives rise
to disconnected branches of the moduli space \rsagnotti.

To summarize, the supergravity point lies on the $B=0$ branch
whereas the Gepner point lies in the $B=1/2$ branch of the $G_2$ moduli
space. Thus it is no surprise that the SCFT computation and the
supergravity computation yield different results.

Does this branch with $B=1/2$ belong to the M-theory moduli space
in four dimensions, as well? The answer is no. In the 
three dimensional M-theory compactification 
on  the $G_2$-manifold times a circle the NS-NS two form flux on the
$G_2$-manifold is lifted to a three form flux, where two components of
$C_{ijk}$ 
lie on the seven dimensional $G_2$ manifold and one component on the
circle. Therefore, decompactifying this model to four dimensions
by making the circle very big, one looses the non-trivial three form 
flux. This is consistent with the result derived in \rwittenb,  that
in M-theory on a Calabi-Yau manifold the non-geometric phases are absent. 

In the case of the quintic the geometric picture is not clear because
we do not understand how to handle the singularities. But for free
involutions the quotient is a genuine manifold and the discrete moduli
must be visible at large volume. And indeed we have seen that
$H_{2}(X;\ZZ)_{\rm tor}=H^3(X;\ZZ)_{\rm tor}\not=\emptyset$, at least
if $h^{11}>1$. So there is the
possibility for a flat but nontrivial $B$--field, that is one with a
characteristic class in $H^3(X;\ZZ)$ that is pure torsion. 
With other words there is the possibility of turning on 
discrete $\ZZ_2$ two form flux through the nontrivial $2$--cycles.

\newsec{Conclusions}

In this paper we have presented a  class of SCFTs describing
certain points in the moduli space of $G_2$ compactifications of
Type II strings. These models are given by anti-holomorphic quotients
of $c=9$ Gepner models times a circle. We have identified  
the anti-holomorphic involution in the $c=9$ SCFT simply as the operation
of charge conjugation in each tensor factor.
However, it turned out that the precise action of this charge
conjugation on all states in the Hilbert space, in particular on the
neutral ones, is not that straightforwardly  to determine. 
Instead for certain simple models we were able to distinguish 
the $\sigma^*$ even from the $\sigma^*$ odd states by intelligent 
guesswork. Of course it would be a big advance  to
find the general solution to this pure CFT problem, which involves
determining the representation theory of a ${\cal W}(2,4,6,8,10)$ 
algebra. This would allow
one, using the general construction given in this paper, to compute
quite a number of exactly solvable points on various $G_2$ manifolds.

Here we have considered three particular SCFTs in quite some detail
including a quotient involving the quintic Calabi-Yau.
In all cases considered in turned out that the SCFT result for the
twisted sector massless spectrum is different  from the supergravity 
result. In was pointed out that this is actually no surprise
taking into account that the Gepner model corresponds to a point
in the non-linear sigma model moduli space where a background
two-form flux has been turned on. In the $G_2$ model this
modulus is frozen to discrete values so that the SCFT and
the naive large volume geometry lie on separate branches
of the moduli space. We also pointed out that the Gepner model
branch is absent in the corresponding four dimensional M-theory 
compactification.

Being equipped with a class of exactly solvable SCFTs one can now
move forward and investigate issues like
the behavior of the model under deformations away from the exactly solvable 
point or the generation of a superpotential.
Moreover, one could try using the abstract boundary state 
\refs{\rbound,\rdbranes,\rkdbranes}
formalism to find stable D-branes in the deep stringy regime 
of $G_2$ compactifications. 

On the one hand one might consider simple generalizations
of the construction presented in this paper, but on the other hand  
one might also contemplate completely different
conformal field theory constructions. which are
 not of the form of a toroidal orbifold
or an anti-holomorphic quotient of a Calabi-Yau times a circle. 
Analogous to \reguchi\ one might start with tensor products of the
${\cal N}=1$ unitary models from the very beginning, even though it
seems to be quite hard to implement a spectral flow in such models.

It is known that  Gepner models are special points in the Landau-Ginzburg phase
of the corresponding $(2,2)$ linear sigma model. 
Thus, it would be interesting to see
whether one can treat these anti-holomorphic quotients directly
in the Landau-Ginzburg model.
It would also be interesting to generalize the construction of 
 superconformal field
theories described in this paper to the case of eight
dimensional manifolds with Spin$(7)$ holonomy. 

\vskip 1cm

\centerline{{\bf Acknowledgments}}\pano
We would like to thank Anamaria Font, Boris K\"ors and Dieter L\"ust
for discussion and Radu Tatar for comments about the manuscript. 
The group is supported in part by the EEC contract ERBFMRXCT96-0045. 
\vfill\eject
\appendix{A}{Unitary HWR of the ${\cal N}=0$ Virasoro algebra}
We summarize some basis data about the ${\cal N}=0$ super Virasoro algebra.
The central charge of the unitary, rational models is
\eqn\censusy{  c=1-{6\over m(m+1)}, \quad
                m=3,4,5,\ldots}
and the HWR are 
\eqn\high{   h^m_{r,s}={ \left( (m+1)r-ms\right)^2-1\over 4m(m+1) } 
               \quad 1\le r\le m-1, 1\le s\le m.}
Moreover, one has the reflection symmetry $h^m_{r,s}=h^m_{m-r,m+1-s}$.
The characters in the NS and R sector can be expressed as
\eqn\chara{\eqalign{
           \chi_{r,s}&={1\over \eta(\tau)}
             \left(
               \Theta_{(m+1)r-ms,m(m+1)}\left({\tau}\right)-
               \Theta_{(m+1)r+ms,m(m+1)}\left({\tau}\right) \right). \cr}}
For the modular S-matrix one gets
\eqn\modulnsiu{
          S_{r_1,s_1;r_2,s_2}=\sqrt{8\over m(m+1)} (-1)^{(r_1+s_1)(r_2+s_2)}
                     \sin\left({\pi r_1 r_2 \over m+1} \right)
                 \sin\left({\pi s_1 s_2 \over m+1} \right) .}
For $m=5$ one obtains the conformal
grid shown in Table 1.
\vskip 0.8cm
\vbox{
\centerline{\vbox{
\hbox{\vbox{\offinterlineskip
\def\tablespace{height2pt&\omit&&\omit&&\omit&&
 \omit&\cr}
\def\tablerule{\tablespace\noalign{\hrule}\tablespace}

\hrule\halign{&\vrule#&\strut\hskip0.2cm\hfill #\hfill\hskip0.2cm\cr
& $3$  && ${7\over 5}$ && ${2\over 5}$ && $0$  &\cr
\tablerule
& ${13\over 8}$  && ${21\over 40}$ && ${1\over 40}$ && ${1\over 8}$ &\cr
\tablerule
& ${2\over 3}$  && ${1\over 15}$ && ${1\over 15}$ && ${2\over 3}$ &\cr
\tablerule
& ${1\over 8}$  && ${1\over 40}$ && ${21\over 40}$ && ${13\over 8}$   &\cr
\tablerule
& $0$  && ${2\over 5}$ && ${7\over 5}$ &&  ${3}$  &\cr
}\hrule}}}}
\centerline{
\hbox{{\bf Table1:}{\it ~~ conformal grid for $m=5$}}}
}
\vskip 0.5cm

\vfill\eject
\appendix{B}{Unitary HWR of the ${\cal N}=1$ super Virasoro algebra}
We summarize some basis data about the ${\cal N}=1$ super Virasoro algebra.
The central charge of the unitary, rational models is
\eqn\censusy{  c={3\over 2}\left(1-{8\over m(m+2)}\right), \quad
                m=3,4,5,\ldots}
and the HWR are 
\eqn\high{   h^m_{r,s}={ \left( (m+2)r-ms\right)^2-4\over 8m(m+2) } 
               +{1-(-1)^{r-s} \over 32},\quad 1\le r\le m-1, 
            1\le s\le m+1}
where $r+s=$even is the NS sector and $r+s=$odd the R-sector.
Moreover, one has the reflection symmetry $h^m_{r,s}=h^m_{m-r,m+2-s}$.
The characters in the NS and R sector can be expressed as
\eqn\chara{\eqalign{
           \chi^{NS}_{r,s}&={1\over \eta(\tau)}
             \sqrt{\theta_3(\tau)\over \eta(\tau)} \left(
               \Theta_{(m+2)r-ms,m(m+2)}\left({\tau\over 2}\right)-
               \Theta_{(m+2)r+ms,m(m+2)}\left({\tau\over 2}\right) \right) \cr
           \chi^{R}_{r,s}&=\left(2-\delta_{r,{m\over 2}}
     \delta_{s,{m+2\over 2}} \right)
           {1\over \eta(\tau)}
             \sqrt{\theta_2(\tau)\over 2\eta(\tau)} \left(
               \Theta_{(m+2)r-ms,m(m+2)}\left({\tau\over 2}\right)-
           \Theta_{(m+2)r+ms,m(m+2)}\left({\tau\over 2}\right) \right).\cr}}
For the modular S-matrix one gets \reholzer\
\eqn\modul{\eqalign{
          S^{NS,NS}_{r_1,s_1;r_2,s_2}&={2\over \sqrt{m(m+2)} }\left(
                 \cos\left({2\pi \lambda_1\lambda_2\over 4 m(m+2)}\right)-
        \cos\left({2\pi \lambda_1\o\lambda_2\over 4 m(m+2)}\right)\right) \cr
      S^{\widetilde{NS},R}_{r_1,s_1;r_2,s_2}&={2\over \sqrt{2m(m+2)} }\left(
                 \cos\left({2\pi \lambda_1\lambda_2\over 4 m(m+2)}\right)-
                     (-1)^{r_1 s_1}
      \cos\left({2\pi \o\lambda_1\lambda_2\over 4 m(m+2)}\right)\right) \cr}}
with $\lambda_i=(m+2)r_i-ms_i$ and $\o\lambda_i=(m+2)r_i+ms_i$.
The conformal for the $m=6$ unitary model is shown in Table 2. 
\vskip 0.8cm
\vbox{
\centerline{\vbox{
\hbox{\vbox{\offinterlineskip
\def\tablespace{height2pt&\omit&&\omit&&\omit&&\omit&&
 \omit&\cr}
\def\tablerule{\tablespace\noalign{\hrule}\tablespace}

\hrule\halign{&\vrule#&\strut\hskip0.2cm\hfill #\hfill\hskip0.2cm\cr
& $3$  && ${29\over 16}$ && ${5\over 6}$ && ${5\over 16}$ && $0$  &\cr
\tablerule
& ${67\over 32}$  && ${33\over 32}$ && ${41\over 96}$ && ${1\over 32}$ && 
 ${3\over 32}$  &\cr
\tablerule
& ${5\over 4}$  && ${9\over 16}$ && ${1\over 12}$ && ${1\over 16}$ && 
 ${1\over 4}$  &\cr
\tablerule
& ${23\over 32}$  && ${5\over 32}$ && ${5\over 96}$ && ${5\over 32}$ && 
 ${23\over 32}$  &\cr
\tablerule
& ${1\over 4}$  && ${1\over 16}$ && ${1\over 12}$ && ${9\over 16}$ && 
 ${5\over 4}$  &\cr
\tablerule
& ${3\over 32}$  && ${1\over 32}$ && ${41\over 96}$ && ${33\over 32}$ && 
 ${67\over 32}$  &\cr
\tablerule
& $0$  && ${5\over 16}$ && ${5\over 6}$ && ${29\over 16}$ && 
 ${3}$  &\cr
}\hrule}}}}
\centerline{
\hbox{{\bf Table 2:}{\it ~~ superconformal grid for $m=6$}}}
}

\listrefs
\end
\bye
\end